\documentclass{achemso}
\usepackage{achemso}
\usepackage{graphicx}
\usepackage{amsmath,amsfonts,amssymb}
\usepackage[version=3]{mhchem}
\usepackage{xspace}
\usepackage{scalerel}

\makeatletter
\newcommand{\whitehourglass}{\mathbin{\scalerel*{\@hgpic}{t}}}
\newcommand{\@hgpic}{%
    \setlength{\unitlength}{0.3cm}
    \begin{picture}(1,1.5)%
    \thicklines%
    \put(0,0){\line(2,3){1}}%
    \put(1,1.5){\line(-1,0){1}}%
    \put(0,1.5){\line(2,-3){1}}%
    \put(1,0){\line(-1,0){1}}%
    \end{picture}%
}
\makeatother

\makeatletter
\newcommand{\latinphrase}[1]{\textit{#1}}

\newcommand{\ie}{\latinphrase{i.e.}\xspace}
\newcommand{\eg}{\latinphrase{e.g.}\xspace}

\makeatother

\title{A comparison of partial atomic charges for electronically excited states}

\author{Ryan J. MacDonell}
\affiliation{Department of Chemistry and Biomolecular Sciences, University of Ottawa, Ottawa, Ontario, K1N 6N5, Canada.}
\altaffiliation{Present address: School of Chemistry, University of Sydney, NSW 2006, Australia}
\author{Serguei Patchkovskii}
\affiliation{Max-Born-Institut f\"{u}r nichtlineare Optik und Kurzzeitspektroskopie, Max-Born-Stra{\ss}e 2A, 12489 Berlin, Germany.}
\author{Michael S. Schuurman}
\email{michael.schuurman@nrc-cnrc.gc.ca}
\affiliation{Department of Chemistry and Biomolecular Sciences, University of Ottawa, Ottawa, Ontario, K1N 6N5, Canada.}
\alsoaffiliation{National Research Council of Canada, 100 Sussex Drive, Ottawa, Ontario K1A 0R6, Canada.}

\date{\today}
\begin{document}
\maketitle

\begin{abstract}
Partial atomic charges are a useful and intuitive concept for understanding molecular properties and chemical reaction mechanisms, showing how
changes in molecular geometry can affect the flow of electronic charge within a molecule. However, the use of partial atomic
charges remains relatively uncommon in the characterization of excited-state electronic structure. Here we show how well-established partial atomic charge methods
perform for interatomic, intermolecular and inter-bond electron transfer in electronically excited states. Our results demonstrate
the utility of real-space partial atomic charges for interpreting the electronic structures that arise in excited-state processes.
Furthermore, we show how this analysis can be used to demonstrate that analogous electronic structures arise near photochemically relevant
conical intersection regions for several conjugated polyenes. Based on our analysis, we find that charges computed using the iterative Hirshfeld
approach provide results which are consistent with chemical intuition and are transferable between homologous molecular systems.
\end{abstract}

\section{Introduction}
Many widely-used chemical concepts lack a rigorous definition as an observable quantum mechanical quantity. Thus, the numerical
quantities that serve as the metrics for these concepts lend themselves to multiple methods of computation which are evaluated primarily
for their utility, rather than ``accuracy''. Examples of such concepts include fundamental aspects of chemistry such as partial atomic charge,
bond order and steric repulsion.\cite{gonthier12} The determination of partial atomic charges
has garnered a significant amount of attention in computational chemistry, since they can be used to quantify shifts in electron density such
as inter- and intramolecular charge transfer, changes in electronic character and potential long-range electrostatic interactions.\cite{gross02,jacquemin12}

Approaches for defining partial atomic charges can be broadly divided into two categories. The first are those which are fit to external
properties such as dipoles, higher multipoles,\cite{cioslowski89,ferenczy91} or the electrostatic potential. An example of the latter is the ``charges
from electrostatic potentials'' (CHELP) family of methods (\eg grid-based, CHELPG),\cite{chirlian87,breneman90} which are useful for
quantifying intermolecular interactions or long-range intramolecular interactions. These methods have the formal advantage of being
defined so as to reproduce a well-defined quantum mechanical property, but may also suffer from ``screening'' particular
atomic sites within large molecules.

The second broad class of methods directly partitions the molecular electron density into atomic components.
It is this class of approaches that will be the focus of the present work.
The preeminent example of this family of methods is Mulliken population analysis, where the partitioning is performed by assigning atomic basis functions to each atomic
centre and determining
\begin{equation}
    \rho_\alpha = \mathrm{Tr}_\alpha (\mathbf{D}\mathbf{S}),
\end{equation}
where $\rho_\alpha$ is the density assigned to atom $\alpha$, $\mathrm{Tr}_\alpha$ is the trace over basis functions centred at atom $\alpha$, $\mathbf{D}$
is the one-electron reduced density matrix and $\mathbf{S}$ is the atomic basis function overlap matrix. The atomic charge is then defined as the difference
between electron density and atomic charge $Z_\alpha$, $q_\alpha = Z_\alpha - \rho_\alpha$. This approach does not readily account for mismatched
basis sets on different atom centres, and diffuse basis functions may give rise to ``spurious'' contributions to atoms which are spatially
separated from the actual electron
density.\cite{mulliken55} Other methods, such as L\"{o}wdin populations or natural atomic orbitals, provide a more robust determination of partial
charges but still depend on assignment based on basis function centres.\cite{lowdin50,reed85} However, partial charges can also be determined
without a knowledge of the underlying atomic basis functions, as for intrinsic atomic orbital populations (for single-reference
wavefunctions),\cite{knizia13,knizia15} quasi-atomic orbitals\cite{west15,west17} and Stockholder projector
analysis.\cite{vanfleteren10,vanfleteren12} These methods project atomic weights onto the full
basis of the molecule, thereby removing the dependence on the atomic centres of the basis set.

Real-space density approaches, in contrast, rely on assigning atomic weights on a grid and integrating the density corresponding to each atom. As a result,
they depend uniquely on the electron density and thus the basis set dependence arises only indirectly, based on the convergence of this quantity.
Population-based approaches are of the form
\begin{equation}
    q_\alpha = Z_\alpha - \int \mathrm{d}\mathbf{r}\, w_\alpha(\mathbf{r}) \rho_\mathrm{mol}(\mathbf{r}),
\end{equation}
where $w_\alpha(\mathbf{r})$ is the weight function (normalized such that $\sum_\alpha w_\alpha(\mathbf{r}) = 1$) and $\rho_\mathrm{mol}(\mathbf{r})$ is the
molecular electron density at cartesian coordinate $\mathbf{r}$. The total electron density is integrated over defined regions of space to give the atomic
populations, and charges are defined as the difference between nuclear charges and populations.
Alternatively, ``promolecular'' approaches rely on densities of spherically-symmetric, non-interacting atoms (the promolecule), \ie
$\rho_\mathrm{pro}(\mathbf{r}) = \sum_\alpha \rho_\mathrm{at}^\alpha(\mathbf{r})$. The difference between promolecular and molecular densities is known as the
deformation density, yielding charges of the form
\begin{equation} \label{eq:pro}
    q_\alpha = \int \mathrm{d}\mathbf{r}\, w_\alpha(\mathbf{r}) \left[ \rho_\mathrm{pro}(\mathbf{r}) - \rho_\mathrm{mol}(\mathbf{r}) \right].
\end{equation}
Real-space approaches to calculating partial charges can thus be differentiated by: (\textit{i}) the representation of atomic charge
(point charges or promolecular densities) and (\textit{ii}) the form of $w_\alpha(\mathbf{r})$. In Bader's quantum theory of atoms in
molecules (QTAIM), 3D space is partitioned according to the electron density
topography, with weight functions of 0 or 1 depending on the atom reached by steepest-ascent of the density.\cite{bader90} Alternatively, Becke charges
use the same formalism as Becke's spherical grids, with size-adjusted Voronoi cells and a sigmoid functional form of $w_\alpha(\mathbf{r})$ between
atomic sites.\cite{becke88} The Hirshfeld approach uses Equation \ref{eq:pro} with a weight function of
$w_\alpha(\mathbf{r}) = \rho_\mathrm{at}^\alpha(\mathbf{r}) / \rho_\mathrm{pro}(\mathbf{r})$, \ie it assigns atomic weights based on contributions to the promolecular
densities.\cite{hirshfeld77} A simpler form, which yields similar results to Hirshfeld charges, is the Voronoi deformation density (VDD) approach,
which assigns a weight of unity to the nearest atomic site of each grid point, thus dividing 3D space into Voronoi cells. These deformation density
approaches generally use neutral, ground-state atomic densities to form the promolecule, and the magnitude of the atomic charges tend to be significantly lower than
other approaches.\cite{fonsecaguerra04} An extension to the Hirshfeld approach treats $q_\alpha$ in Equation \ref{eq:pro} as a change in charge,
$\delta q_\alpha$, and uses weighted sums of charged atomic densities, iterating until a self-consistent charges are determined, which are generally
independent of the starting conditions. This method, known as iterative Hirshfeld (IH), effectively ``fits'' atomic densities to the molecular density
and tends to yield charges in good agreement with CHELPG and QTAIM.\cite{bultinck07}

Although each of the partial atomic charge approaches listed above have been extensively used for the analysis of molecular ground electronic states, there are
comparatively far fewer examples for which these techniques have been used for the analysis of excited-state wavefunctions. Calculating consistent values
of charge requires converged electron densities, which themselves require
additional computational effort for the more complex, multi-reference electronic characters of molecules in excited states. Nonetheless, the charge
methods discussed above are equally applicable for excited state properties. For example, QTAIM analysis has been used to evaluate charges and bonding
interactions of electronically excited formaldehyde.\cite{ferrocostas14} A range of charge methods were assessed for charge-transfer
excitations of a series of conjugated push-pull organic compounds, and showed that electrostatic potential methods best reproduce differences
in dipole moments for vertical excitations.\cite{jacquemin12} Other approaches have been used for bonding analyses\cite{jaracortes17} and to quantify
non-covalent interactions\cite{vannay16} for excited molecules.

Another example is excited-state charge transfer, which is a characteristic of many photochemical processes. It is thus frequently the subject of experimental
and theoretical study.\cite{rodriguezmayorga15,chiu13,aquino14} Many heteronuclear diatomics have well-studied avoided crossings between covalent and ion-pair
states, where dramatic changes in the atomic charges are realized as the inter-nuclear separation increases and are evinced by the dipole
moment.\cite{varandas09,rodriguezmayorga15,jasik17} Molecular
complexes may also have charge transfer states if the differences in electron affinity between molecules is sufficiently high. This results in the
loss of an electron from the donor molecule into an unoccupied (or partially occupied) orbital of an acceptor molecule, creating an ion pair. Some
examples of this are tetracyanoethylene (TCNE) with tetramethylethylene (TME),\cite{li02} benzene and polycyclic aromatic
hydrocarbons.\cite{chiu13,aquino14,vannay16} The ethylene $\pi$-bond of TCNE acts as an electron acceptor, and excitation of the complexed molecule
gives the transition TCNE-X $\rightarrow$ TCNE$^{-}$-X$^{+}$ which leads to a stronger intermolecular bond in the excited state.\cite{aquino14,vannay16}

Charge transfer can also play a role over shorter distances in the ultrafast photochemistry of molecules. One well-known example is the ``sudden polarization''
of ethylene: following excitation to the $\pi\pi$* state, ethylene may undergo internal conversion to the ground state by a torsion about the \ce{C=C}
bond followed by a pyramidalization of a single methylene (\ce{CH2}) group to reach a minimum energy conical intersection (MECI) between the two states, shown in
Scheme \ref{sch:ethy} (with the $\whitehourglass$ symbol representing conical intersections, R = R$'$ = H, and labels \textbf{C$N$Pyr} represent
pyramidalization at carbon $N$). The electronic character on the excited state resembles
that of a lone pair on the pyramidalized carbon, resulting in a significant dipole moment across the \ce{C=C} bond.\cite{wulfman71,bonacickoutecky75,brooks79}
This pathway is not unique to ethylene: it is common to substituted ethylenes (R or R$'$ $\neq$ H in Scheme \ref{sch:ethy}),\cite{macdonell18,herperger20} as well as
many larger alkenes such as butadiene,\cite{olivucci93,norton06,levine09,glover18} substituted
butadienes,\cite{olivucci94,macdonell19,macdonell20} hexatriene and cyclohexadiene\cite{norton06,wolf19} and larger biomolecules such as the
retinal chromophore,\cite{grobner00,polli10,nogly18} all of which may undergo photochemical isomerization by Scheme \ref{sch:ethy} to yield a photoproduct
\textbf{2} different from the starting molecule \textbf{1}. Polyenes with two
or more conjugated \ce{C=C} bonds also have a low-lying ($\pi$*)$^2$ state which can undergo internal conversion through nonpolar conical
intersections (\eg ``kinked-diene'' MECIs such as ``transoid'', labelled \textbf{C$N$Tr} for displacement about carbon $N$).\cite{garavelli06} Thus,
understanding the branching ratios of excited-state decay can be aided by characterizing local changes in electron
density.\cite{macdonell18,macdonell19,macdonell20,herperger20}

\begin{scheme}
    \centering
    \includegraphics[width=\textwidth]{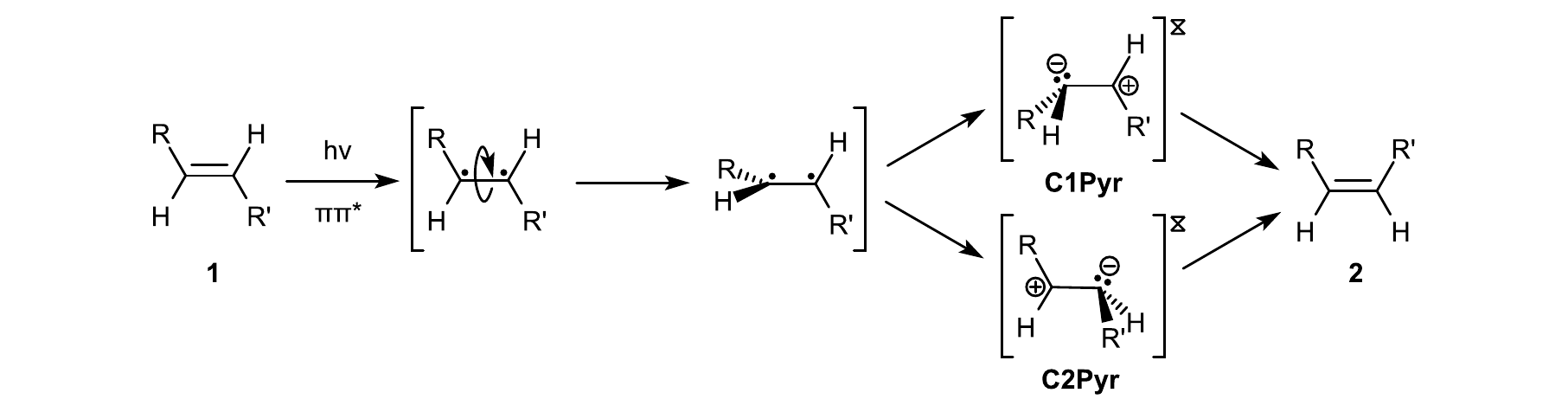}
    \caption{}
    \label{sch:ethy}
\end{scheme}

It should be noted that the adiabatic electronic states are not uniquely defined at a precise point of degeneracy (\ie a conical intersection).
However, to elucidate the coupled nuclear-electronic dynamics that define the mechanisms of ultrafast photochemistry, it is often desirable to
study the properties of the excited electronic state on the \textit{approach} to the conical intersection region. At the point of
degeneracy the adiabatic states are inter-mixed, but can be resolved by projecting onto a geometry on the decay pathway prior to the conical intersection,
or by making an infinitesimal displacement along a branching space coordinate.\cite{atchity91,yarkony97} In Section \ref{sec:poly},
we use the latter approach to study trends in charges of characteristic MECIs of polyenes.

In this paper, we show how a variety of partial atomic charge methods perform for characterization of excited-state wavefunctions. We focus on examples
where the charge-transfer character is well known: interatomic charge-transfer of LiH and LiF, intermolecular charge transfer of the benzene-TCNE (B-TCNE)
complex and the sudden polarization effect of ethylene, 1,3-butadiene, 1,3,5-hexatriene, 1,3-cyclohexadiene and their derivatives substituted with
amino ($\pi$-donor) and cyano ($\pi$-acceptor) groups. We show how partial atomic charges can be used as a coarse-grained view of the otherwise complex
electronic characters of excited states, and how similarities in charge differences may plan an important role in identifying motifs characteristic of
photochemistry.

\section{Computational methods}
Electron densities and their gradients were calculated at each point of a Becke molecular grid\cite{becke88} with Lebedev spherical
quadrature.\cite{lebedev75,lebedev76,lebedev99}
Densities and calculated charges were converged with a small number of radial and spherical points. All calculations employ
70 radial and 110 angular points, \ie 770 grid points per atom. Density values were calculated from the one-electron reduced density matrix (1-RDM)
or, equivalently, from natural molecular orbitals and their respective populations. QTAIM charges and properties were calculated using Multiwfn 3.7 with
a grid spacing of 0.06 Bohr for all molecules unless otherwise stated.\cite{multiwfn,tang09}

Ground-state atomic densities for the deformation density promolecule were calculated for atoms with charges of $-3$ to $+3$ using restricted open-shell
second order M{\o}ller-Plesset perturbation theory (ROMP2) with an aug-cc-pVTZ basis in the GAMESS electronic structure package.\cite{knowles91,gamess}
Details of the symmetrization of atomic densities are given in Section 1 in the supporting information (SI).
These promolecule densities were
used in the calculation of Hirshfeld, IH and VDD charges. Becke charges were calculated using the same weighting
procedure as the Becke grids. Thus, the points of each spherical grid were assigned to the nucleus at its centre.\cite{lehtola14}

A series of test calculations were also performed to assess the charges and their basis set dependence. The charges for several small molecular
systems are shown in Figure S1 in the SI, where the basis set label indicates the basis used to calculate the electron density of
the molecule. For comparison, Mulliken charges are shown with the expected divergence as the size of the atomic basis increases. As previously reported,
Hirshfeld and VDD charges closely resemble each other and have nearly the same value as the Mulliken cc-pVDZ charges.\cite{fonsecaguerra04} Becke
charges are also similar to Hirshfeld and VDD with slightly greater magnitudes for most molecules. Iterative Hirshfeld (IH) has the same sign as the other
grid-based charges,
but is much greater for all cases except carbon atom of HCN. This greater magnitude is consistent with QTAIM and CHELPG results,\cite{bultinck07} although
QTAIM shows differences for HCN and formaldehyde by assigning significantly more charge to the C and O atoms, respectively. The
only strong basis set dependence of the grid-based methods occurs for formaldehyde (\ce{OCH2}), which is due to an unconverged electron density
of the molecule with the smaller cc-pVDZ and cc-pVTZ bases.

Since the purpose of the present study is to compare the utility of different partial atomic charge schemes, we are satisfied to simply
choose a level of theory in the computation of the electronic wave function that is sufficient for producing reasonably accurate electron densities
for each specific molecule. Furthermore, all comparisons between methods will be made with the same electron density.
Ground- and excited-state calculations of LiH and LiF were performed using single and double excitation multi-reference configuration interaction (MR-CISD)
and multi-state complete active space second-order perturbation theory (MS-CASPT2)\cite{finley98} with a
Dunning augmented triple-zeta basis (aug-cc-pVTZ) with a 2-electron, 5-orbital complete active space averaged over three complete active space
self-consistent field (CASSCF) reference states (SA3-2,5-CAS) for LiH, and a SA2-6,6-CAS reference for LiF. B-TCNE geometries were optimized using
resolution-of-the-identity M{\o}ller-Plesset perturbation theory (RI-MP2) with a double-zeta basis (cc-pVDZ) using the Q-Chem electronic structure
package,\cite{qchem} and excited-state calculations were performed with CASPT2 with an augmented double-zeta basis (aug-cc-pVDZ) using a SA4-4,4-CAS
reference. All MR-CISD and (MS-)CASPT2 calculations were performed using the Molcas electronic structure package.\cite{molcas8a,molcas8b} Minimum-energy
conical intersections (MECIs) were optimized and excited-state properties of polyenes were
calculated with multi-reference single-excitation configuration interaction (MR-CIS) and a 6-31G* basis using the COLUMBUS electronic structure
package.\cite{columbus} Due to the low-lying $\pi$3s Rydberg states of ethylene and vinylamine, a diffuse s-type basis function was added to the centre of
mass of ethylene and the nitrogen atom of vinylamine, both generated with the \texttt{genano} routine in Molcas.\cite{molcas8a,molcas8b} Ground- and
excited-state minimum energy geometries and minimum energy conical intersections (MECIs)
were optimized at the same level of theory. Additional details on the level of theory for each molecule are given in Section 9 in the SI.

\section{Interatomic charge transfer}
\begin{figure}
    \centering
    \includegraphics[width=0.5\textwidth]{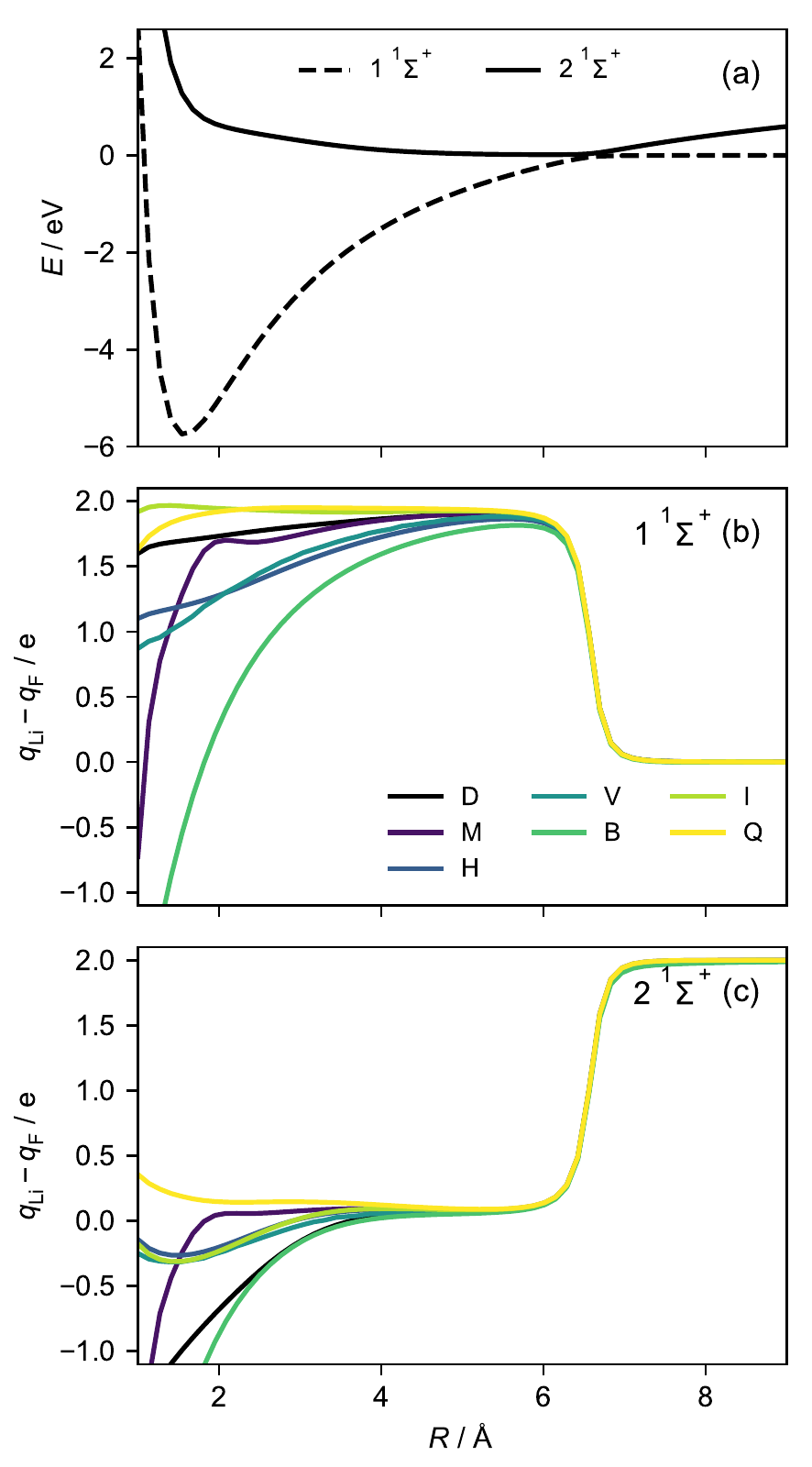}
    \caption{(a) MR-CISD potential energy curves of LiF and (b--c) charge differences as a function of bond length for each state. Charges in (b--c) are D: dipole;
    M: Mulliken; H: Hirshfeld; V: VDD; B: Becke; I: IH; and Q: QTAIM, with colour legend shown in panel (b).}
    \label{fig:lif}
\end{figure}

One of the simplest examples of charge transfer in chemistry are the covalent and charge-transfer states of heteronuclear diatomics. Figure \ref{fig:lif}a
shows the lowest two singlet states of LiF with $\Sigma^+$ symmetry as a function of the bond length. There is a narrowly avoided crossing at an
interatomic distance of 6.6 \r{A} with $\Delta E = 0.07$ eV. From a Lewis picture, the ground state has an ionic \ce{Li^+-F^-} character whereas the
first excited state is covalent,
\ce{Li-F}. At the dissociation limit, the ground state is made up of neutral atoms (Li + F) and the excited state is the corresponding atoms (Li$^+$ +
F$^-$).\cite{varandas09} The partial atomic charges in Figure \ref{fig:lif}b show suggest a more complex evolution of the electronic character. For
comparison, charges derived from the dipole moment ($q_\mathrm{D} = 2\mu/R$ for dipole moment with magnitude $\mu$) are also shown. At bond
lengths shorter than equilibrium, there are two qualitatively different behaviours: Mulliken and Becke charges are negative for both states (\ce{Li^--F^+}
for $\Delta q = q_\mathbf{Li} - q_\mathbf{F} = -2$) with a similar divergence to the 2 $^1\Sigma^+$ dipole charge, whereas Hirshfeld, VDD, IH and QTAIM charges
are positive for 1 $^1\Sigma^+$ and close to zero for
2 $^1\Sigma^+$. In the vicinity of the equilibrium bond length, only IH and QTAIM match the Lewis picture with $\Delta q = 2$, whereas Hirshfeld, VDD
and Mulliken charges have $\Delta q \approx 1$ on the ground state. All methods have a crossing in partial charges at the avoided crossing.
At bond lengths greater than 7 \r{A}, all charge methods match the neutral and ionic characters expected.
Of these methods, Mulliken, IH and QTAIM are closest to the expected charge difference from the ground-state equilibrium to greater bond length, with the Mulliken
charges diverging for shorter bond lengths.

For comparison, Figure S2 in the SI shows potential energies and charge differences of LiF using CASSCF, CASPT2 and MS-CASPT2. CASSCF is well known
to predict an avoided crossing at a much shorter bond length of 4.2 \r{A}. Adding perturbative corrections leads to an unphysical crossing of the adiabatic states,
but the corresponding perturbed density (seen by the charges) is barely changed. On first glance, MS-CASPT2 appears to correct the crossing leading to accurate
potential energies. Charges were calculated using perturbationally-modified CASSCF,\cite{finley98} \ie the CASSCF wavefunctions transformed by the MS-CASPT2
diagonalization, which reveals that at short bond lengths the CASSCF electronic character remains with an intermediate peak in the charge difference at the CASSCF
avoided crossing point. With known accurate electronic properties of LiF,\cite{varandas09} this example demonstrate how the charges reveal errors in the electronic
character where the potentials do not.

\begin{figure}
    \centering
    \includegraphics[width=0.5\textwidth]{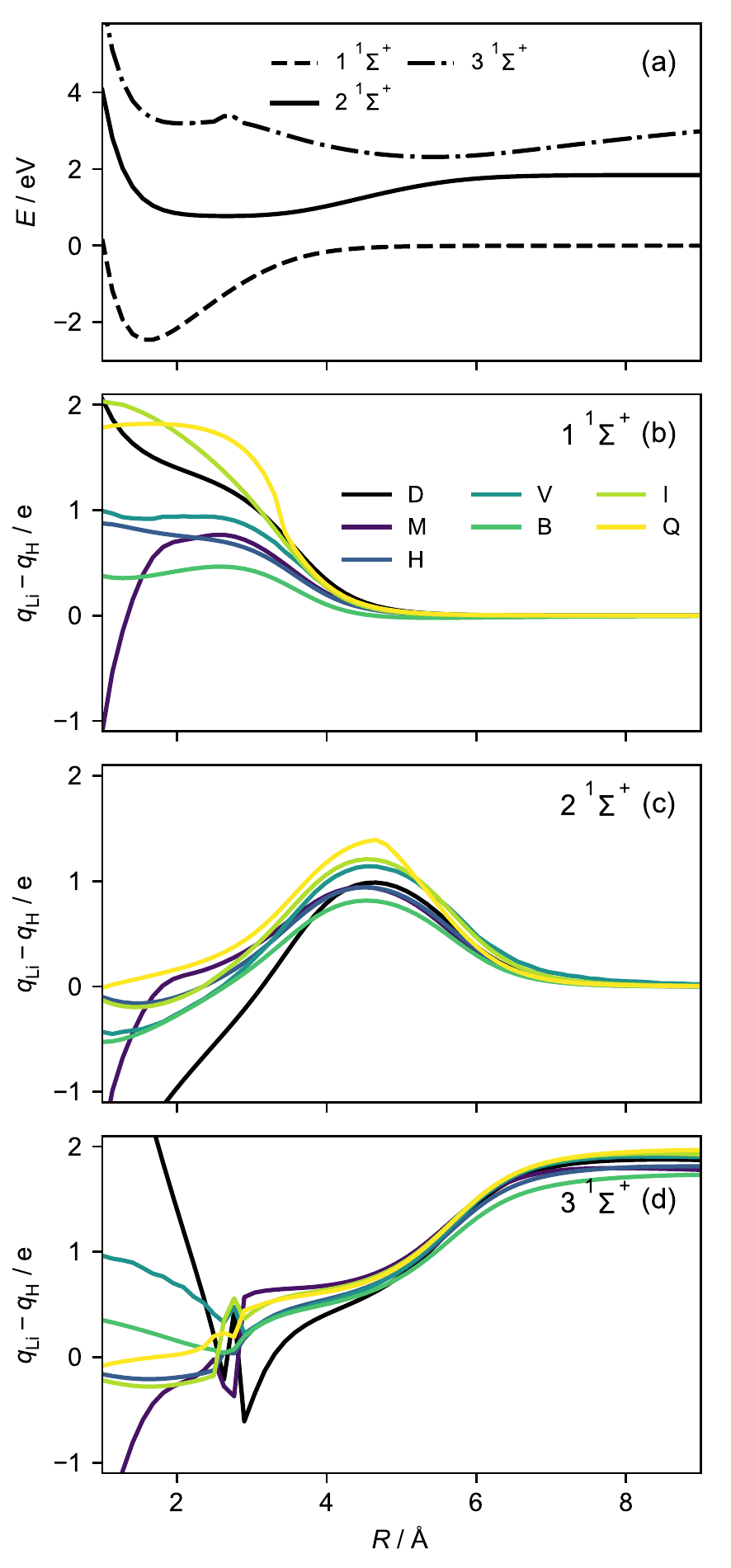}
    \caption{(a) MR-CISD potential energy curves of LiH and (b--d) charge differences as a function of bond length for each state. Charges in
    (b--d) are D: dipole; M: Mulliken; H: Hirshfeld; V: VDD; B: Becke; I: IH; and Q: QTAIM, with colour legend shown in panel (b).}
    \label{fig:lih}
\end{figure}

A more complex example is given by LiH. The lowest $^1\Sigma^+$ state of LiH is ionic, much like LiF. The two lowest $^1\Sigma^+$ excited states
of LiH both correspond to neutral \ce{Li-H}. Approaching the dissociation limit, the ground adiabatic state corresponds
to the ground-state ions Li(2s) + H(1s), the first excited state involves an excitation of Li to give Li(2p) + H(1s) and S$_2$ is the ionic state, Li(1s)$^+$ +
H(1s)$^-$. At longer bond lengths, the second excited state re-crosses a higher dissociative state.\cite{jasik17} The ionic state thus passes through both neutral
states during dissociation. Figure \ref{fig:lih} shows potential energies and charge differences for these three states as a function of bond length. The avoided
crossing around 3--4 \r{A} shows strong coupling between the adiabatic states, with a minimum energy difference of 1.2 eV. In Figure \ref{fig:lih}b--d,
the partial atomic charges show (\textit{i}) very different behaviours at short bond lengths, (\textit{ii}) similar charges with different magnitudes
between the two avoided crossings, and (\textit{iii}) nearly identical values approaching the dissociation limit.

At bond lengths shorter than equilibrium, Mulliken gives negative charge differences for all three states, whereas VDD and Hirshfeld show similar behaviours on
the ground state ($\Delta q \approx 0.9$), and Becke ($\Delta q = 0.4$) is qualitatively different from IH and QTAIM ($\Delta q \approx 2.0$).
On the first excited state, Becke and VDD have similar magnitudes near $-0.5$, and Hirshfeld and IH are also similar near $-0.1$ whereas QTAIM has $\Delta q = 0$.
The state characters change in the vicinity of the first avoided crossing to yield maximum charges on the first excited state
for all methods, and aside from Hirshfeld and VDD the methods over-estimate the maximum relative to the dipole charge. Charges for the second excited state with
all grid-based methods differ significantly at short bond lengths, particularly VDD which predicts a charge difference of 1 and the dipole charge which diverges
in the direction of positive values. QTAIM has a non-nuclear attractor in the vicinity of the 3 $^1\Sigma^+$ avoided crossing at 2.5 \r{A}, the density of which was
added to Li to avoid a sharp discontinuity. All methods are roughly in agreement by the avoided crossing with a value of $\sim$1
before reaching values of 1.7--2 at the dissocation limit. As in the case of LiF, IH and QTAIM charges exhibit the same trends as each other for all electronic states.
LiH exhibits relatively little dynamic electron correlation, as seen in Figure S3 in the SI, where the potentials and charges are roughly the same
with CASSCF, CASPT2 and MS-CASPT2. The PM-CASSCF charges exhibit small peaks not seen in the MS-CASPT2 potentials, which may suggest a greater sensitivity to
intruder states.

The trends in charges give some indication of the suitability of methods for different analyses: all methods give qualitatively similar results at longer
bond lengths. Mulliken and Becke charges do not match chemical intuition at short bond lengths where the orbital overlap is highest for Mulliken and the
Becke weight function is most significant near the atom centres. Hirshfeld and VDD schemes yield charges whose sign matches expected values, but whose
magnitude is small. IH and QTAIM give the expected signs and magnitudes of charges for most of the dissociation path in both LiF and LiH when comparing to
dipole charges, although they do not exhibit divergence at short bond lengths. Some of these
results, particularly the small magnitudes of Hirshfeld and VDD,\cite{bultinck07} have been noted previously for small molecules.

\section{Intermolecular charge transfer}
\begin{figure}
    \centering
    \includegraphics[width=0.5\textwidth]{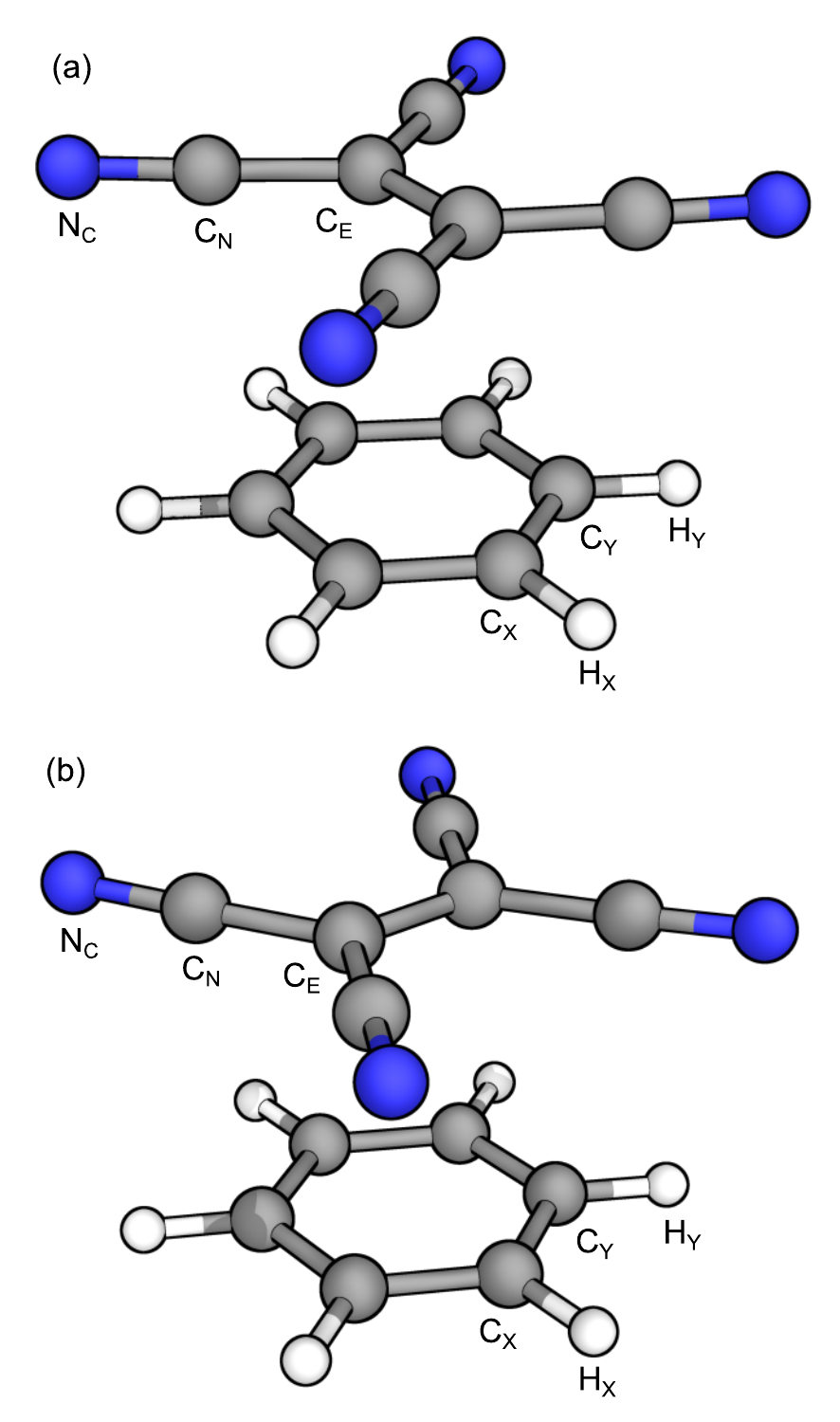}
    \caption{Structures of the B-TCNE complex in (a) parallel and (b) perpendicular geometries. Labels for symmetry-equivalent atoms are shown for
    both.}
    \label{fig:btstruct}
\end{figure}

Charge transfer states in molecular complexes correspond to the excitation of an electron from an occupied orbital in one molecule to an unoccupied or
partially occupied orbital in another. These occur in $\pi$-stacked donor-acceptor complexes, where the acceptor has a higher electron affinity and
can take an electron from the donor. An example of this is the benzene-tetracyanoethylene (B-TCNE) complex, with TCNE acting as the acceptor and benzene
as the donor. The B-TCNE complex has two low energy structures with C$_\mathrm{2v}$ symmetry with nearly identical energies. Figure \ref{fig:btstruct}
shows the two structures, denoted ``parallel'' (par.) and ``perpendicular'' (perp.), as well as labelled symmetry-equivalent atoms. The centres of
mass of the two molecules are separated by a distance of 3.1 \r{A} at the minimum energy point.

Table \ref{tab:btcne} shows the energies and differences in molecular charge ($\Delta q_{\text{B-T}}$) for the ground state and the three lowest-lying
excited states: two charge-transfer states and a higher-lying $\pi$3s Rydberg state. At the CASPT2/aug-cc-pVDZ level of theory, the S$_0$ energy of the perpedicular
geometry is only 4 meV above that of the parallel geometry. The charge differences show a significant variation for different charge
methods considering the distance between the two molecules. Hirshfeld, VDD, IH and QTAIM charges are in good agreement for all states of both geometries, and
are for the most part smaller in magnitude than Mulliken and Becke charges. All methods suggest a $\Delta q_{\text{B-T}}$ of roughly 2 e for the first excited state
(S$_1$) and a smaller value for the second excited state (S$_2$). The greatest disparity appears for S$_3$,
in which an electron from TCNE is excited into a 3s Rydberg orbital with a centre
closer to TCNE. Relative to the Hirshfeld, VDD, IH and QTAIM values, the Mulliken charges are appear underestimated whereas the Becke charges are overestimated.
For Mulliken, this is readily explained by its tendency to assign electron density to all atom centres with diffuse basis functions despite the large spatial
distance separating the atomic centres. Becke partitioning appears to have the opposite effect: the Voronoi weighting assigns more of the density to TCNE than benzene
relative to the other approaches (aside from VDD). Interestingly, the VDD charges of the neutral states of B-TCNE appear to more closely resemble IH and QTAIM despite
the difference in weighting factors.

\begin{table}
    \caption{CASPT2 potential energies ($E$) and molecular charge differences ($\Delta q_{\text{B-T}}$) for the ground and excited states of The B-TCNE complex.}
    \label{tab:btcne}
    \begin{tabular*}{\textwidth}{l@{\extracolsep{\fill}}cccccccc}
        \hline \hline
        & & & \multicolumn{6}{c}{$\Delta q_{\text{B-T}}$ / e} \\ \cline{4-9}
        Geometry~ & State & Energy / eV & Mulliken & Hirshfeld & VDD & Becke & IH & QTAIM \\ \hline
        par. & S$_0$ & 0.000 & $0.28$ & $0.15$ & $0.18$ & $0.31$ & $0.09$ & $0.20$ \\
        & S$_1$ & 3.437 & $1.82$ & $1.58$ & $1.63$ & $1.76$ & $1.67$ & $1.76$ \\
        & S$_2$ & 3.476 & $2.03$ & $1.77$ & $1.83$ & $1.96$ & $1.87$ & $1.96$ \\
        & S$_3$ & 6.958 & $0.11$ & $0.34$ & $0.34$ & $0.61$ & $0.29$ & $0.40$ \\ \hline
        perp. & S$_0$ & 0.004 & $0.37$ & $0.16$ & $0.19$ & $0.32$ & $0.11$ & $0.21$ \\
        & S$_1$ & 3.445 & $1.88$ & $1.58$ & $1.64$ & $1.76$ & $1.68$ & $1.75$ \\
        & S$_2$ & 3.471 & $2.09$ & $1.78$ & $1.84$ & $1.96$ & $1.88$ & $1.96$ \\
        & S$_3$ & 6.760 & $0.29$ & $0.44$ & $0.44$ & $0.69$ & $0.43$ & $0.50$ \\
        \hline \hline
    \end{tabular*}
\end{table}

\begin{figure}
    \centering
    \includegraphics[width=0.73\textwidth]{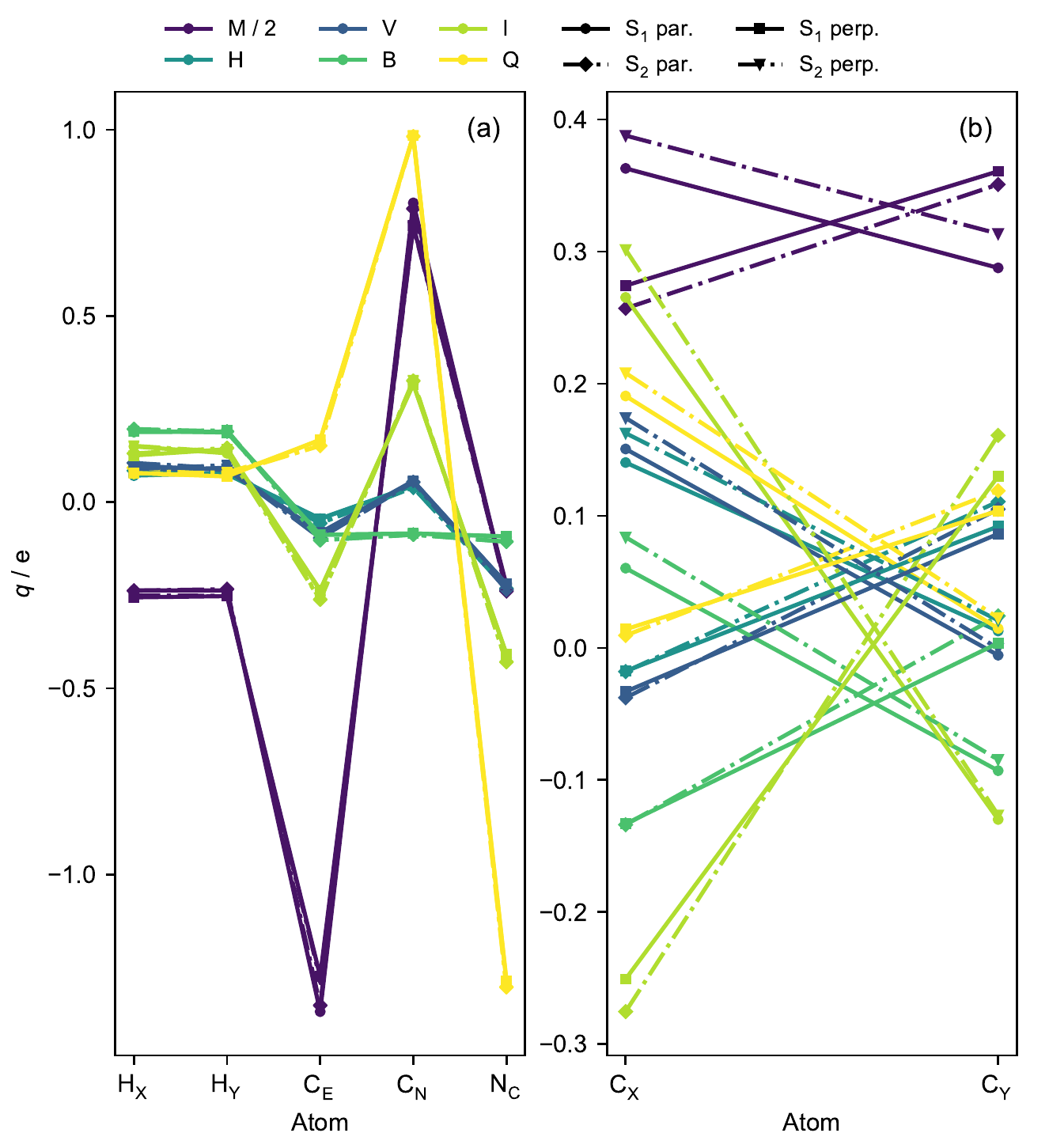}
    \caption{CASPT2 charges of symmetry-equivalent (a) benzene H atoms and all TCNE atoms (b) benzene C atoms for the charge-transfer states of B-TCNE. Charge labels
    are M: Mulliken; H: Hirshfeld; V: VDD; B: Becke; I: IH; and Q: QTAIM. Mulliken charges were scaled by a factor of 2 to fit on the same scale.}
    \label{fig:btcchg}
\end{figure}

Examining solely the molecular charge differences allows us to identify the two charge-transfer states of the B-TCNE complex, but does little to
distinguish the electronic structure between them. The two charge-transfer states result from excitation of an electron from nearly-degenerate benzene $\pi$-orbitals
(E$_\mathrm{1g}$ symmetry for D$_\mathrm{6h}$ benzene) to the empty $\pi$*-orbital of TCNE. The presence of TCNE breaks the benzene orbital symmetry,
leading to one orbital with lobes at the C$_\mathrm{X}$ atoms, and the other with nodes in the same positions. This is observed by the charges of equivalent
atoms in Figure \ref{fig:btcchg} which shows values for S$_1$ and S$_2$. The Mulliken approach predicts charges much greater in magnitude than the other
methods, and thus had to be scaled by a factor of 2. QTAIM assigns a much greater charge to the atoms in the TCNE cyano group, C$_\mathrm{N}$ and
N$_\mathrm{C}$, with a difference greater than 2 e across the \ce{C#N} bond. All other approaches are roughly in agreement with smaller differences in magnitude.
The difference between states can be seen particularly from the C$_\mathrm{X}$ and C$_\mathrm{Y}$ charges in Figure \ref{fig:btcchg}b, which swap values from
S$_1$ to S$_2$. Comparing parallel and perpendicular geometries reveals that the charges also swap from the parallel to the perpendicular structure. Importantly, this
demonstrates how partial atomic charges may be used to assign the electronic character by identifying chemically relevant differences in the electron density.
To quantify the effect of dynamic electron correlation on these results, Table S3 and Figure S4 in the SI show charges calculated using
the CASSCF wavefunctions. The trends for all methods are identical, but the two states swap energetic order and the magnitudes increase slightly in the absence
of dynamic correlation.

\section{Photodynamics of polyenes} \label{sec:poly}
\begin{figure}
    \centering
    \includegraphics[width=0.5\textwidth]{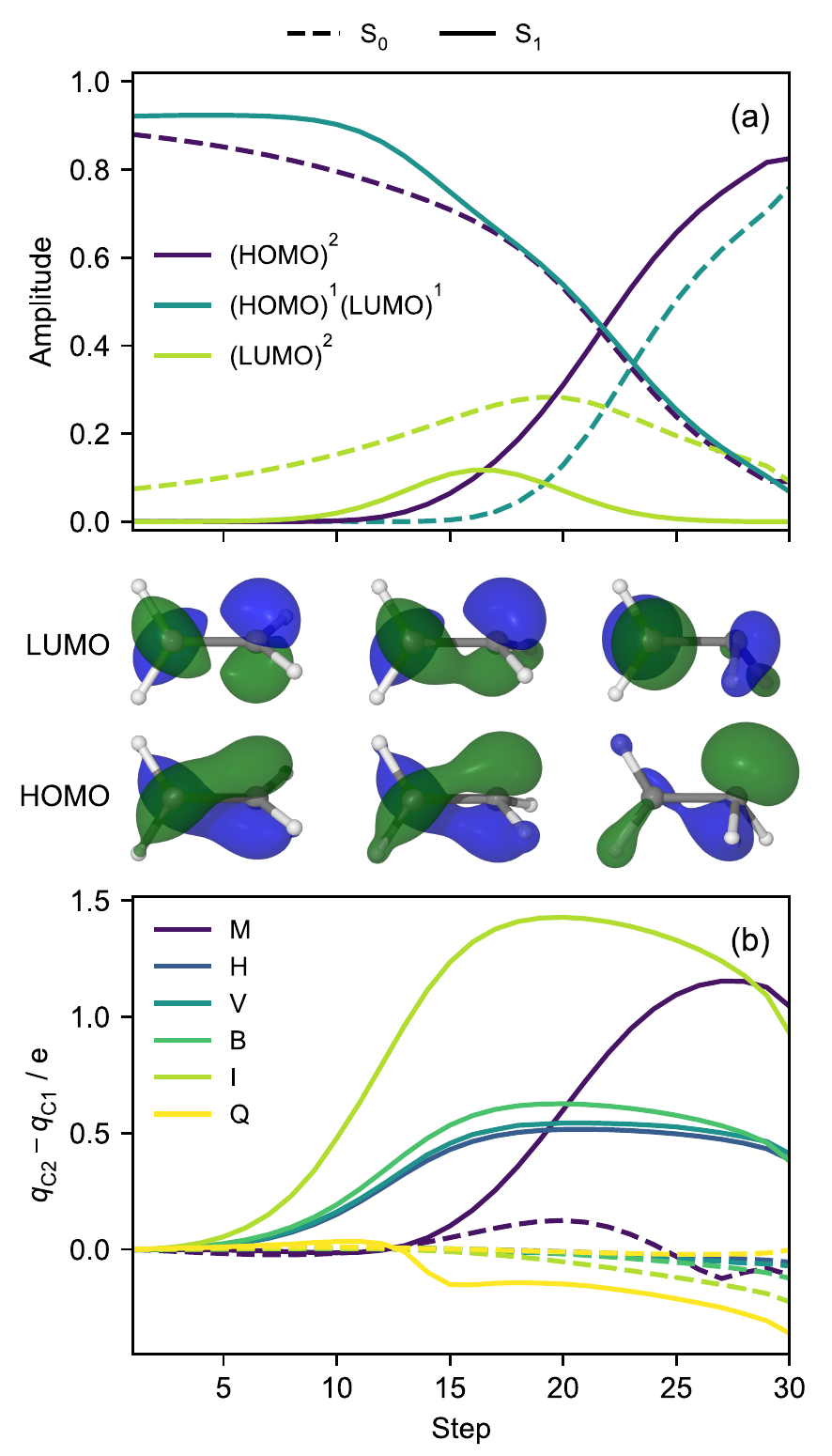}
    \caption{Evolution of (a) electronic character and (b) charge differences along interpolated internal coordinates from the S$_2$-S$_1$ twisted MECI to
    the \textbf{C1Pyr} S$_1$-S$_0$ MECI. The HOMO and LUMO are shown at representative points of the path. Charges in (b) are M: Mulliken; H: Hirshfeld;
    V: VDD; B: Becke; I: IH; and Q: QTAIM. Dashed lines represent the ground state (S$_0$) and solid lines represent the excited state (S$_1$) in (a) and (b).}
    \label{fig:sudpol}
\end{figure}

The characteristic nuclear motion of the excited state dynamics of ethylene (following excitation to the bright $\pi\pi$* state) is a twisting of
the \ce{C=C} bond followed by the pyramidalization of one methylene (\ce{CH2}) group to reach the \textbf{C1Pyr} twist-pyramidalization
MECI (Scheme~\ref{sch:ethy}). Concurrent with this geometric change is the sudden polarization across the \ce{C=C} bond, which involves a shift of
electron density to the pyramidalized site. This is illustrated by the evolution from the twisted (symmetric)
S$_2$-S$_1$ MECI to the \textbf{C1Pyr} MECI, shown in Figure \ref{fig:sudpol}. The geometries between MECIs were generated by interpolating along natural
internal coordinates,\cite{pulay92} which mainly consisted of the out-of-plane pyramidalization motion, \ce{CH2} scissoring and the methylene tilt. The pyramidalization
and scissoring motion have been shown before to most strongly modulate the coupling between the $\pi\pi$* state and ground state,\cite{viel03} whereas the tilt has a
relatively minor effect lowering the potential energy and reducing the molecule to C$_1$ symmetry overall. Without the tilt, there is a point on the
conical intersection seam only 0.3 eV bove the MECI with C$_s$ symmetry which bridges between equivalent MECI geometries.\cite{mori13} As the nuclear
geometry evolves, so too does the electronic character. Figure \ref{fig:sudpol}a shows the dominant configurations of the MR-CIS wavefunction as a
function of the geometry for the ground and excited state, along with the ``highest-occupied'' and ``lowest-unoccupied'' CASSCF molecular orbitals (HOMO and LUMO,
respectively, assuming a closed-shell ground-state configuration). The potential energies of all calculated states are shown in Figure S5 in
the SI for reference. Initially, the ground state (dashed lines) is mostly in a closed-shell $\pi^2$
configuration whereas the excited state (solid lines) is in the open-shell $\pi\pi$* configuration. As pyramidalization occurs, the symmetry between
carbons is broken and the configurations swap, with a brief mixing of the doubly-excited ($\pi$*)$^2$ character at intermediate geometries. The excited-state
character is mostly (HOMO)$^2$ at the MECI, which resembles a lone-pair orbital at the pyramidalized carbon.

Figure \ref{fig:sudpol}b shows the change in atomic charge difference across the \ce{C=C} bond as a function of the geometry using several methods. Here, C1
represents the carbon atom at which pyramidalization takes place. Aside from QTAIM, each of the methods show qualitatively similar behaviour of pyramidalization
leading to polarization across the \ce{C=C} bond on S$_1$, with more negative charge at C1. Each method exhibits a maximum charge diference on S$_1$
before reaching the \textbf{C1Pyr} conical intersection region. In this respect, Mulliken charges disagree with other methods with the onset of polarization
and the maximum value occuring much later along the interpolated path. Becke charges yield similar results to the promolecular methods (Hirshfeld and VDD)
with maximum charge differences around 0.5 e, whereas IH yields maximum charge differences around 1.4 e around the same geometry, with a decrease
to 1.0 e at the \textbf{C1Pyr} MECI.

The QTAIM charges of the S$_1$ state of ethylene show the opposite of the expected behaviour, tending towards a positive charge on the pyramidalized carbon with
a local minimum halfway through pyramidalization. These charges were more sensitive to grid size than past examples, requiring a grid separation of 0.04 Bohr for
convergence. To understand the behaviour of QTAIM, we first looked at the charges using the reference CASSCF wavefunctions, shown in
Figure S6 in the SI. Although all of the CASSCF charges (including Mulliken and QTAIM) obey the same dominant trend from Figure \ref{fig:sudpol}b,
they are greater in magnitude by nearly a factor of two. The difference between MR-CIS and CASSCF QTAIM results is shown in Figure S7 in the SI: at the
MR-CIS level of theory, the separatrix between C1 and C2 QTAIM basins approaches C1 in a way that counteracts the expected charge difference, whereas the CASSCF
separatrix remains at the centre of the \ce{C=C} bond. The presence of the dummy atom did not play a significant role, with an identical trend along the same path
with the dummy atom removed from the calculation; however, the dummy atom does have significant effect on IH charges in Figure S6b. At the CASSCF
level of theory, S$_1$ has a dominant Rydberg character at the start of the path. As the molecular begins to polarize, the Rydberg character leads to a steep
gradient in the IH charge iterations that cannot be ``fit'' to any proatomic density. Increasing the grid size had no effect on the calculated IH charges, suggesting
IH may have an underlying instability when basis functions not located at atom centres have a significant population.

Also shown in Figure \ref{fig:sudpol}b are the differences in ground-state charges. With the exception of Mulliken
charges, each method undergoes a gradual decrease in charge with the minima at the S$_1$-S$_0$ MECI. Mulliken charge differences on S$_0$ initially
decrease, then increase in the region of the S$_1$ maxima for mother methods, then decrease to a minimum value before the MECI. The difference between
Mulliken charges and other methods may in part be from the Rydberg 3s basis function located between carbon atoms, to which Mulliken populations may be
assigned.

The sudden polarization effect and twist-pyramidalization conical intersection motif is not unique to ethylene. It is in fact common for larger polyenes
to have similar geometric motifs to the \textbf{C1Pyr} MECI of ethylene (\textbf{C$N$Pyr} for pyramidalization at carbon $N$), while also
having competitive (and dominant, for longer chain lengths) MECIs
which are nonpolar and have a greater multireference character. These kinked-diene (\textbf{C$N$Tr}) MECIs have mixed electronic structures which
have tetraradical character. In the case of the ethylenic MECIs, the valence orbitals are ``localized'' about the
pyramidalized group and the potential energy is largely invariant with chain length. This localization also appears for other systems such as silicon
nanoparticles, and may be a general trait of conical intersections due to the relatively small energetic cost of achieving degeneracy between the localized
HOMO and LUMO at symmetry-broken geometries.\cite{levine19a,levine19b} The kinked-diene MECIs are more delocalized than the ethylenic MECIs, and do not
exhibit the sudden polarization effect.

\begin{figure}
    \centering
    \includegraphics[width=0.5\textwidth]{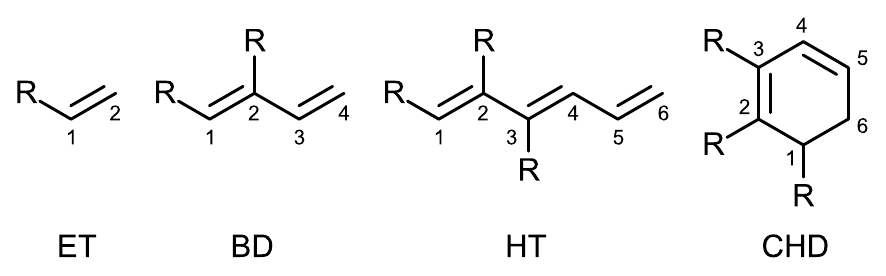}
    \caption{Carbon numbering and possible substituent positions for the polyenes ET, BD, HT and CHD. Indices of CHD were chosen to match HT.}
    \label{fig:pstruct}
\end{figure}

\begin{figure}
    \centering
    \includegraphics[width=0.73\textwidth]{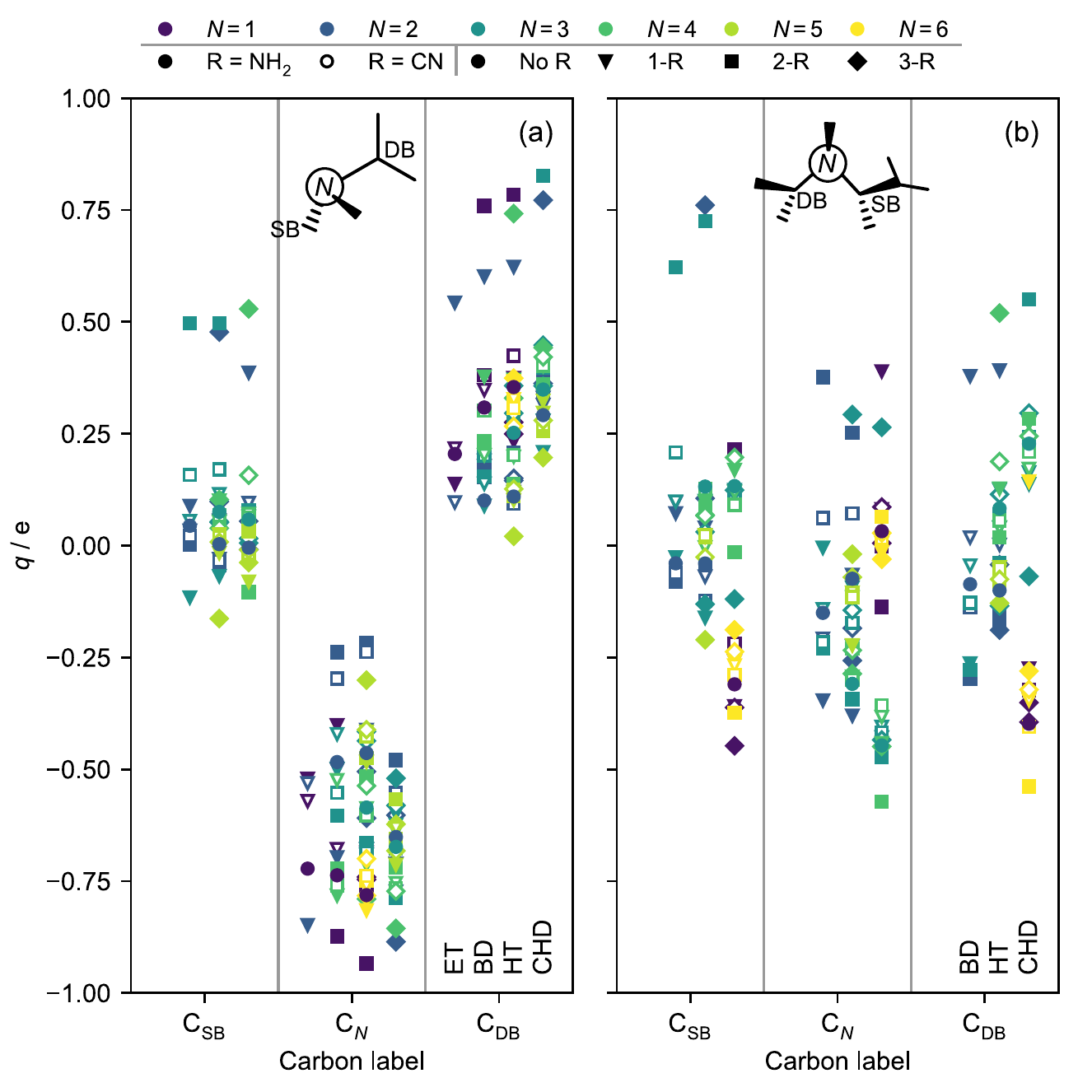}
    \caption{Iterative Hirshfeld charges of polyene MECI geometries at pyramidalized ($\mathrm{C}_N$) sites, and sites doubly-bonded ($\mathrm{C_{DB}}$)
    and singly-bonded ($\mathrm{C_{SB}}$) to $\mathrm{C}_N$. Ethylenic geometries are
    shown in (a), and kinked-diene geometries are given in (b). Colours represent pyramidalization position ($N$), shapes represent
    substituent position and filled/open shapes represent the amino/cyano substituent identity. From left to right in each section is ET, BD, HT and CHD.
    Polyene backbone structures for the two types of MECIs are shown inset.}
    \label{fig:polyci}
\end{figure}

To demonstrate the similar electronic structures of polyenes conical intersections, a set of 148 MECIs of ethylene (ET), 1,3-butadiene (BD),
1,3,5-hexatriene (HT), 1,3-cyclohexadiene (CHD) and their amino (Am) and cyano (CN) singly-substituted derivatives were optimized at the MR-CIS/6-31G*
level of theory.  Their labeling used in the text is given in Figure \ref{fig:pstruct}. Each geometry was displaced by 1 pm along the
gradient-difference vector in order to break the MECI degeneracy so that the state characters could be assigned. If the two state characters remained
unresolved, the displacement along the nonadiabatic coupling vector was sufficient to resolve them. Charges on the closed-shell state of ethylenic
geometries and the mixed tetraradical state of kinked-diene MECIs at the pyramidalized (or central) carbon and the double-bonded (DB) and single-bonded
(SB) carbons are shown in Figure \ref{fig:polyci} (note that bond orders correspond to Figure \ref{fig:pstruct}, except for the ring-closure MECIs of CHD
where the C1--C6 bond is DB for the sake of showing both bonded carbons). The state characters were chosen to match that of
ET and BD on S$_1$ approaching the \textbf{C1Pyr} and \textbf{C2Tr} MECIs, respectively. All charges were evaluated using the IH method due to its consistent
behaviour and large magnitudes shown in the previous sections.

The charges in Figure \ref{fig:polyci} are divided into ethylenic and kinked-diene groups (Figure \ref{fig:polyci}a and b, respectively) to show the
differences in local electron density between the two conical intersection types. Most optimized ethylenic MECIs show a strong polarization of the pyramidalized
carbon relative to the adjacent carbons, whereas kinked-diene geometries have values closer to zero with several MECIs showing the opposite trend in
charge. This becomes clearer when comparing the average polarization across the carbon-carbon bonds: for ethylenic geometries,
$\Delta q_{\mathrm{DB}} = q_\mathrm{DB} - q_N = 0.92 \pm 0.29$ and $\Delta q_{\mathrm{SB}} = 0.65 \pm 0.24$, whereas for the kinked-diene geometries,
$\Delta q_{\mathrm{DB}} = 0.12 \pm 0.43$ and $\Delta q_{\mathrm{SB}} = 0.15 \pm 0.40$. The outliers in Figure \ref{fig:polyci} are all amino-substituted
molecules (filled shapes), and all correspond to the site adjacent to the substituted carbon atom. Interestingly, these outliers have more positive charges,
contrary to the partial negative charge expected from resonance of an amino $\pi$-donor.
Also noteworthy are the geometries in \ref{fig:polyci}b with significantly large charge differenes, such as the \textbf{C3Tr}
MECIs of 2-AmBD, 2-AmHT and 2-AmCHD ($N = 3$, filled squares), or polarization in the opposite direction such as the \textbf{C2Tr} MECIs of 2-AmBD and 2-AmHT
($N = 2$, filled squares). It should be noted that the alternation of charge continues down the polyene backbone for most MECIs, much like the resonance
picture.

The variability in charges of polyene and substituted polyene MECIs shows that great care must be taken in using charges alone to assign branching
ratios for nonadiabatic dynamics; however, in previous studies\cite{macdonell18,macdonell19,macdonell20,herperger20} we have shown that differences in
pyramidalization angles in addition to differences in charge often provide a robust separation of different MECIs. Substituents have a minor effect on
the charge at MECIs, but they have a significant effect on the potential energy gradients which direct an excited wavepacket to different MECIs.
Figure S8 in the SI shows the energies of the S$_1$-S$_0$ MECIs relative to the MECI pyramidalized at the carbon favoured by the substituent
(\ie with a relatively large electron density). In all cases, amino-substitution lowers the energy of pyramidalization at the doubly-bonded carbon below that
of all other ethylenic MECIs, and cyano-substitution lower the energy at the substituted site below other ethylenic MECIs aside from 1- and 3-CNHT.
The same trend is not reproduced for the kinked-diene MECIs aside from those of substituted BD and 2-CNHT.
The change in ethylenic MECI energies is readily predicted by the resonance picture of $\pi$-acceptor and $\pi$-donor substituents and their
stabilization of electron density (\ie during pyramidalization) at specific sites. Due to the corresponding changes gradients on the $\pi\pi$*
state,\cite{macdonell18,macdonell19,macdonell20,herperger20} these results help substantiate the claim that functional group substituents can be used to steer
photochemical reactions in a predictable manner.

\section{Conclusions}
Partial atomic charges have been long-employed to characterize the electronic structure of ground-state molecules and rationalize
reaction mechanisms. The fact that these quantities are not quantum mechanical observables has led to the development of numerous definitions and
computational strategies for their calculation. Real-space charges, in particular, can be efficiently implemented on a grid and offer a robust determination
of charge with no explicit dependence on atomic basis sets. The real-space methods that we tested in this study converge with a minimal grid size, aside from QTAIM
where the separatrix must be well-defined by the grid points.\cite{tang09}
As such, there is no formal or computational barrier to their deployment in the description of excited state electronic structures and properties.
Rather, it is simply the intuitive understanding of the strengths and weaknesses of these techniques for the specific bonding motifs that arise in excited state
dynamics that is lacking, due in part to a dearth of computational data.

Molecules representing three classes of excited-state charge migration were presented here: (\textit{i}) curve crossings between covalent and
ion-pair states, (\textit{ii}) inter-molecular charge transfer, and (\textit{iii}) transient polarization associated with a common conical intersection motif.
All methods that we tested readily identify the ion-pair states for large atomic separation, but only IH and QTAIM identify the same character
for bond lengths shorter than equilibrium. All approaches readily identify the differences in B-TCNE charge transfer states and minimum-energy geometries
with the exception of the Mulliken charges. The real-space methods also show consistent onset of sudden polarization of ethylene on the excited state and the
lack of polarization on the ground state, although the polarization direction is reversed for QTAIM with MR-CIS densities. From these results, we have identified
that IH both produces charges consistent with chemical intuition and most clearly differentiates local charge environments with comparatively large partial charge
magnitudes. Finally, we used IH charges to begin to gain an intuition for the photodynamics of polyenes: ET, BD, HT and CHD showed a consistent
trend of local carbon-carbon bond polarization at ethylenic conical intersections, but no trend for kinked-diene conical intersections. These results
point towards a simple picture of polyene excited-state dynamics which can be used to predict and design novel photochemical properties.

\begin{acknowledgement}
    The authors would like to thank Issaka Seidu and Simon P. Neville for many helpful discussions. R.J.M. and M.S.S. acknowledge financial support from the
    Natural Sciences and Engineering Research Council of Canada (NSERC).
\end{acknowledgement}

\bibliography{ms}
\end{document}


\maketitle
\beginsupplement

\section{Spherically averaged atomic densities} \label{sec:atomdens}
Due to the degeneracy of atomic orbitals with like angular momentum $l$, there
are a wide range of asymmetric solutions for the electron density of an atom
which give the same energy. In the absence of interactions to break the
spherical symmetry (as assumed for atomic components of the promolecular
density), the wavefunction is a state or superposition of states that is
spherically symmetric. It can be shown that the radial density of a spherically
symmetric atomic wavefunction is the same as an asymmetric density integrated
over the angular components.

To store ab initio radial densities, we note that the total density is
given by
\begin{equation}
    \rho_{tot} = \mathrm{Tr}(\mathbf{DS}) = \sum_{ij} D_{ij} S_{ij} =
    \sum_{ij} D_{ij} A_{ij} R_{ij},
\end{equation}
where $\mathbf{D}$ is the one-electron reduced density matrix (1-RDM),
$\mathbf{S}$ is the overlap matrix, $\mathbf{A}$ is the angular overlap matrix
and $\mathbf{R}$ is the radial overlap matrix. Given the angular overlaps,
the 1-RDM can be reduced to a purely radial form with only $N_\mathrm{shell} \times
N_\mathrm{shell}$ terms
\begin{align}
    D_{R,ij}^\prime &= D_{ij} A_{ij}, \\
    \mathbf{D}_R &= \mathbf{T}^T \mathbf{D}_R^\prime \mathbf{T}, \label{eq:dr}
\end{align}
where $\mathbf{T}$ is an $N_\mathrm{basis} \times N_\mathrm{shell}$ matrix with elements of 0 and 1,
which sums over all basis functions in the same shell.

A cartesian atomic orbital basis has the form
\begin{equation}
    \psi_i(x, y, z) = N_i x^{a_i} y^{b_i} z^{c_i} \sum_j c_{ij} \exp\left(-\zeta_{ij} (x^2 + y^2 + z^2)\right),
\end{equation}
where $x$, $y$ and $z$ are cartesian distances from the atom centre, $a_i$, $b_i$ and $c_i$ are
their respective exponents, $c_{ij}$ are contraction coefficients and $\zeta_{ij}$ are primitive
exponents of the Gaussian basis. The elements of $\mathbf{A}$ for the cartesian basis are then given by
\begin{equation}
    A_{ij} = N_i N_j \int_0^{2\pi} \mathrm{d}\theta\, \cos^{a_i + a_j} \theta
    \sin^{b_i + b_j} \theta \int_0^{\pi} \mathrm{d}\phi\,
    \sin^{1+a_i+a_j+b_i+b_j} \phi \cos^{c_i + c_j} \phi,
\end{equation}
where $N_i$ and $N_j$ satisfy $A_{ii} = 1$. Integration yields zero if
$x_i + x_j$ is odd for $x \in {a, b, c}$, otherwise
\begin{align}
    A_{ij} &= \frac{\sqrt{(2 l_i + 1)!!(2 l_j + 1)!!}}{(l_i + l_j + 1)!!}
    \prod_x \frac{(x_i + x_j - 1)!!}{\sqrt{(2 x_i - 1)!!(2 x_j - 1)!!}}, \\
    l_i &= \sum_x x_i = a_i + b_i + c_i.
\end{align}
Thus, $\mathbf{A}$ can be used to calculate $\mathbf{D}_R$ from Equation \ref{eq:dr}, and the radial wavefunction
$\psi_l(r)$ and density $\rho(r)$ at radial distance $r = \sqrt{x^2 + y^2 + z^2}$ are given by
\begin{align}
    \psi_l(r) &= \frac{r^l}{\sqrt{2l + 1}} \sum_j c_{lj} \exp(-\zeta_{lj} r^2), \\
    \rho(r) &= \sum_{kl} D_{R,kl} \psi_k (r) \psi_l (r).
\end{align}

\section{Example partial charge calculations}
\begin{figure}[H]
    \centering
    \includegraphics[width=\textwidth]{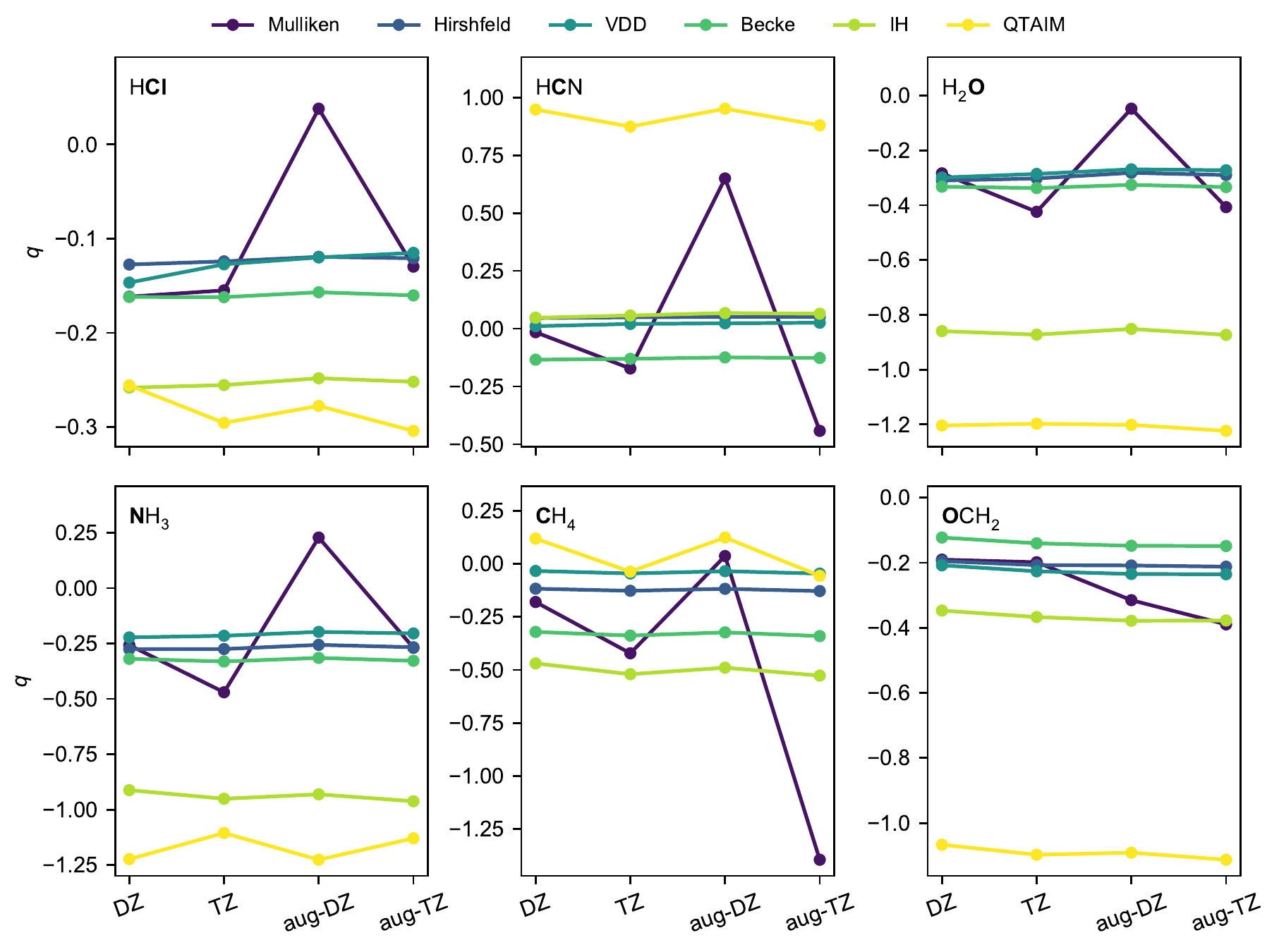}
    \caption{Partial charges of atoms (shown in bold) with increasing basis set size. Basis labels correspond to Dunning basis sets (cc-pV*).}
    \label{fig:exchg}
\end{figure}

\section{Li-X sample charge calculations}
\begin{figure}[H]
    \centering
    \includegraphics[width=\textwidth]{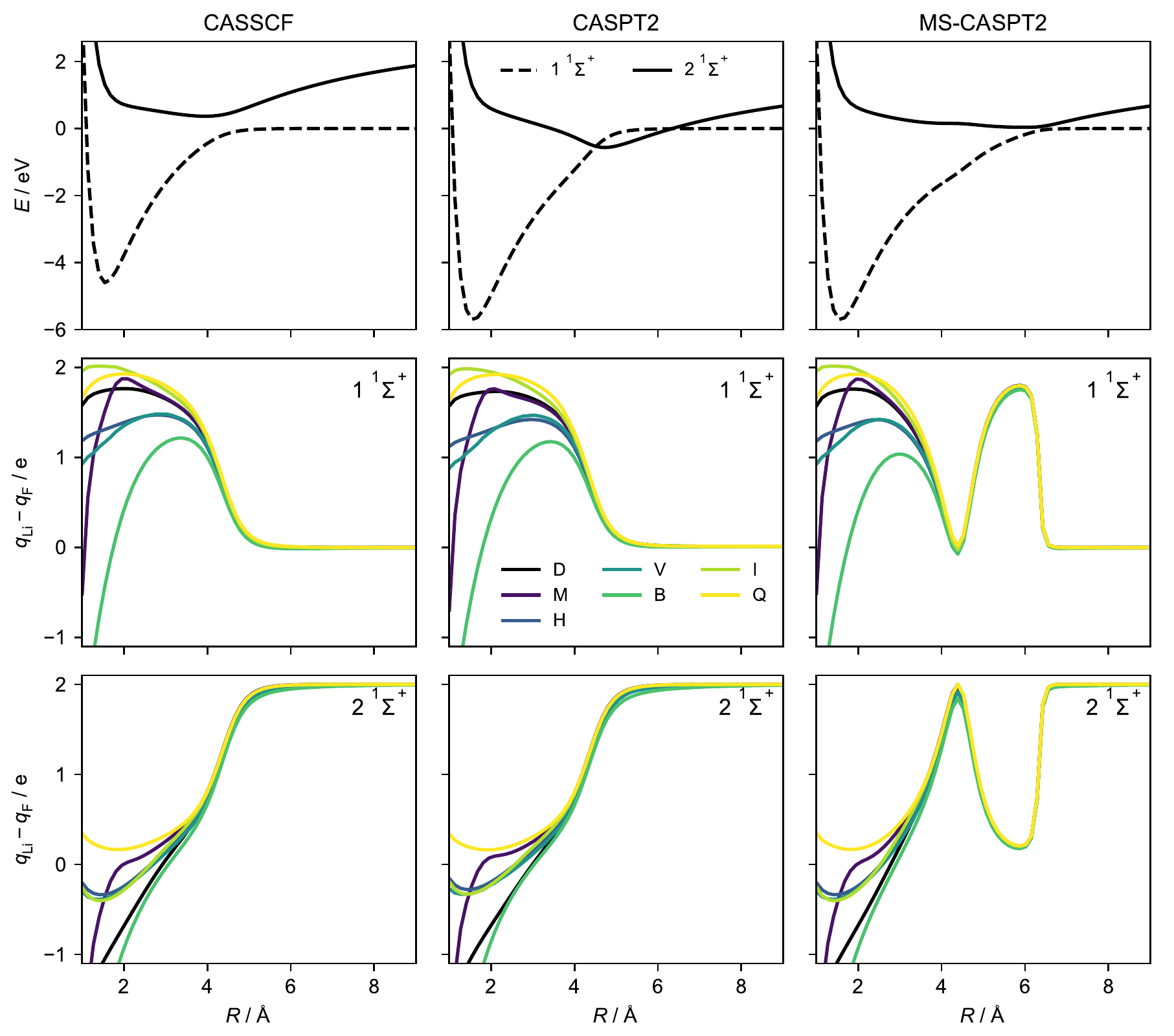}
    \caption{Potential energies (top row) and LiF charge differences (lower rows) as a function of bond length for each state with different electronic
    structure methods (from left to right: CASSCF, CASPT2, MS-CASPT2 with PM-CASSCF charges). Charges are
    D: dipole; M: Mulliken; H: Hirshfeld; V: VDD; B: Becke; I: IH; and Q: QTAIM. Legends apply to all equivalent panels.}
    \label{fig:lifpt2}
\end{figure}

\begin{figure}[H]
    \centering
    \includegraphics[width=\textwidth]{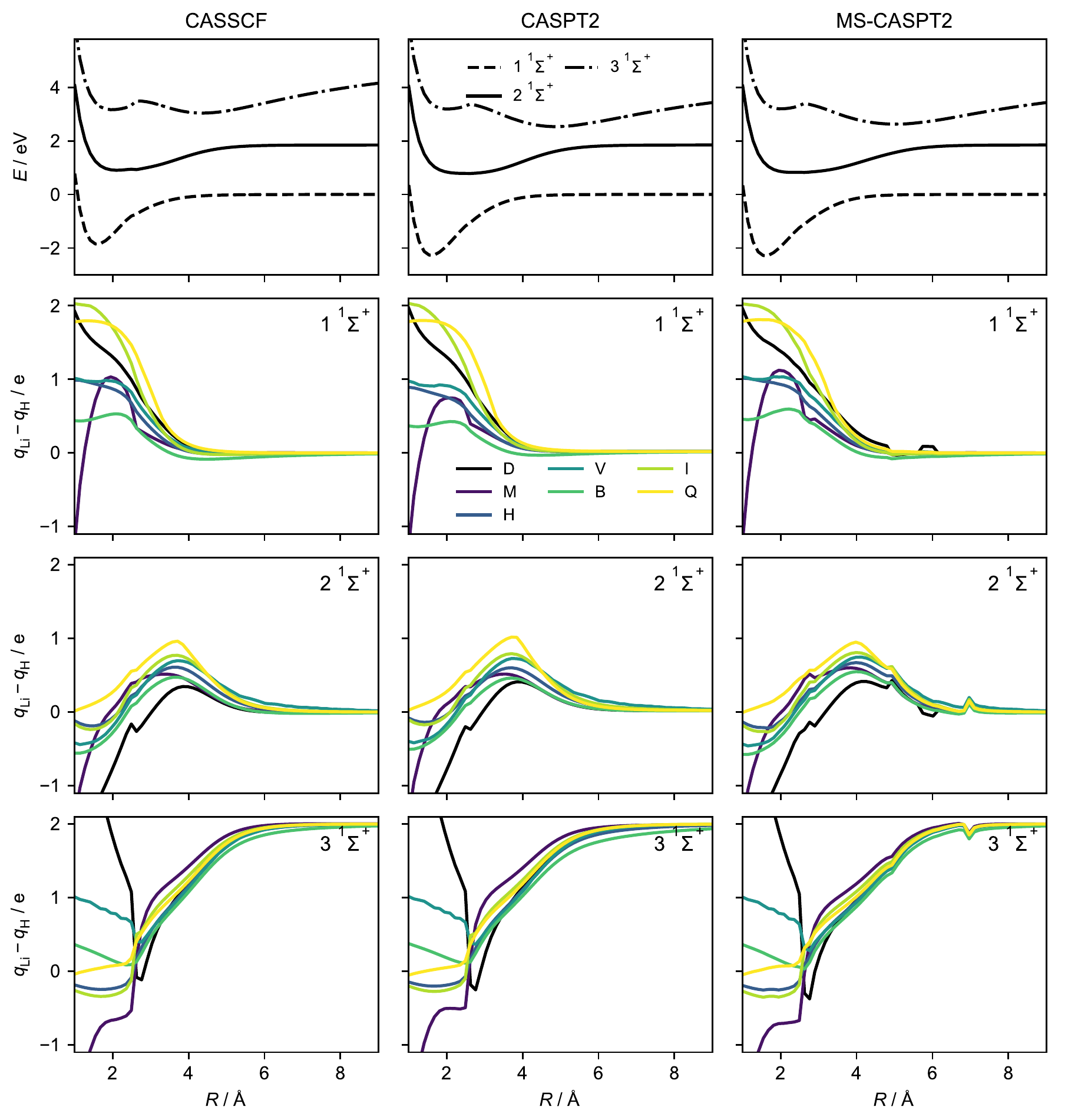}
    \caption{Potential energies (top row) and LiH charge differences (lower rows) as a function of bond length for each state with different electronic
    structure methods (from left to right: CASSCF, CASPT2, MS-CASPT2 with PM-CASSCF charges). Charges are
    D: dipole; M: Mulliken; H: Hirshfeld; V: VDD; B: Becke; I: IH; and Q: QTAIM. Legends apply to all equivalent panels.}
    \label{fig:lihpt2}
\end{figure}

\section{B-TCNE optimized geometries}
\begin{table}[H]
    \centering
    \caption{RI-MP2/cc-pVDZ optimized geometry of the B-TCNE parallel conformer.}
    \begin{tabular*}{0.7\textwidth}{l@{\extracolsep{\fill}}rrr}
        \hline
        C & $ 1.0000000000$ & $ 1.4040633832$ & $ 1.8493301258$ \\
        C & $ 1.2197368544$ & $ 0.7021233104$ & $ 1.8561422595$ \\
        C & $ 1.2197368544$ & $-0.7021233104$ & $ 1.8561422595$ \\
        C & $ 0.0000000000$ & $-1.4040633832$ & $ 1.8493301258$ \\
        C & $-1.2197368544$ & $-0.7021233104$ & $ 1.8561422595$ \\
        C & $-1.2197368544$ & $ 0.7021233104$ & $ 1.8561422595$ \\
        H & $ 2.1660545174$ & $ 1.2513015032$ & $ 1.8577501554$ \\
        H & $ 0.0000000000$ & $ 2.4990849330$ & $ 1.8518337101$ \\
        H & $-2.1660545174$ & $ 1.2513015032$ & $ 1.8577501554$ \\
        H & $-2.1660545174$ & $-1.2513015032$ & $ 1.8577501554$ \\
        H & $ 0.0000000000$ & $-2.4990849330$ & $ 1.8518337101$ \\
        H & $ 2.1660545174$ & $-1.2513015032$ & $ 1.8577501554$ \\
        C & $ 0.0000000000$ & $ 0.6904506291$ & $-1.1697854202$ \\
        C & $ 0.0000000000$ & $-0.6904506291$ & $-1.1697854202$ \\
        C & $ 1.2314008601$ & $ 1.4268422850$ & $-1.2069454010$ \\
        C & $-1.2314008601$ & $ 1.4268422850$ & $-1.2069454010$ \\
        C & $-1.2314008601$ & $-1.4268422850$ & $-1.2069454010$ \\
        C & $ 1.2314008601$ & $-1.4268422850$ & $-1.2069454010$ \\
        N & $-2.2459005156$ & $ 2.0470839090$ & $-1.2611267878$ \\
        N & $-2.2459005156$ & $-2.0470839090$ & $-1.2611267878$ \\
        N & $ 2.2459005156$ & $-2.0470839090$ & $-1.2611267878$ \\
        N & $ 2.2459005156$ & $ 2.0470839090$ & $-1.2611267878$ \\
        \hline
    \end{tabular*}
\end{table}

\begin{table}[H]
    \centering
    \caption{RI-MP2/cc-pVDZ optimized geometry of the B-TCNE perpendicular conformer.}
    \begin{tabular*}{0.7\textwidth}{l@{\extracolsep{\fill}}rrr}
        \hline
        C & $-1.4076730501$ & $ 0.0000000000$ & $-1.8558029762$ \\
        C & $-0.7038691863$ & $-1.2166816375$ & $-1.8573170043$ \\
        C & $ 0.7038691863$ & $-1.2166816375$ & $-1.8573170043$ \\
        C & $ 1.4076730501$ & $ 0.0000000000$ & $-1.8558029762$ \\
        C & $ 0.7038691863$ & $ 1.2166816375$ & $-1.8573170043$ \\
        C & $-0.7038691863$ & $ 1.2166816375$ & $-1.8573170043$ \\
        H & $-1.2524695502$ & $-2.1637363787$ & $-1.8633247472$ \\
        H & $-2.5019811575$ & $ 0.0000000000$ & $-1.8546607995$ \\
        H & $-1.2524695502$ & $ 2.1637363787$ & $-1.8633247472$ \\
        H & $ 1.2524695502$ & $ 2.1637363787$ & $-1.8633247472$ \\
        H & $ 2.5019811575$ & $ 0.0000000000$ & $-1.8546607995$ \\
        H & $ 1.2524695502$ & $-2.1637363787$ & $-1.8633247472$ \\
        C & $ 0.0000000000$ & $ 0.6902938444$ & $ 1.1741729407$ \\
        C & $ 0.0000000000$ & $-0.6902938444$ & $ 1.1741729407$ \\
        C & $-1.2321386545$ & $ 1.4254549796$ & $ 1.2107417227$ \\
        C & $ 1.2321386545$ & $ 1.4254549796$ & $ 1.2107417227$ \\
        C & $ 1.2321386545$ & $-1.4254549796$ & $ 1.2107417227$ \\
        C & $-1.2321386545$ & $-1.4254549796$ & $ 1.2107417227$ \\
        N & $ 2.2481732506$ & $ 2.0431132417$ & $ 1.2665360124$ \\
        N & $ 2.2481732506$ & $-2.0431132417$ & $ 1.2665360124$ \\
        N & $-2.2481732506$ & $-2.0431132417$ & $ 1.2665360124$ \\
        N & $-2.2481732506$ & $ 2.0431132417$ & $ 1.2665360124$ \\
        \hline
    \end{tabular*}
\end{table}

\section{B-TCNE CASSCF charges}
\begin{table}[H]
    \caption{CASSCF potential energies ($E$) and molecular charge differences ($\Delta q_{\text{B-T}}$) for the ground and excited states of The B-TCNE complex.}
    \label{tab:mcbtcne}
    \begin{tabular*}{\textwidth}{l@{\extracolsep{\fill}}cccccccc}
        \hline \hline
        & & & \multicolumn{6}{c}{$\Delta q_{\text{B-T}}$ / e} \\ \cline{4-9}
        Geometry~ & State & Energy / eV & Mulliken & Hirshfeld & VDD & Becke & IH & QTAIM \\ \hline
        par. & S$_0$ & 0.01 & $0.28$ & $0.14$ & $0.17$ & $0.30$ & $0.06$ & $0.22$ \\
        & S$_1$ & 2.67 & $2.08$ & $1.83$ & $1.89$ & $2.02$ & $1.94$ & $2.00$ \\
        & S$_2$ & 2.97 & $1.88$ & $1.64$ & $1.70$ & $1.82$ & $1.73$ & $1.81$ \\
        & S$_3$ & 6.50 & $0.16$ & $0.39$ & $0.38$ & $0.66$ & $0.34$ & $0.46$ \\ \hline
        perp. & S$_0$ & 0.00 & $0.35$ & $0.15$ & $0.19$ & $0.31$ & $0.07$ & $0.22$ \\
        & S$_1$ & 2.78 & $2.14$ & $1.83$ & $1.90$ & $2.01$ & $1.93$ & $2.00$ \\
        & S$_2$ & 2.85 & $1.94$ & $1.64$ & $1.70$ & $1.81$ & $1.74$ & $1.80$ \\
        & S$_3$ & 6.53 & $0.37$ & $0.49$ & $0.50$ & $0.74$ & $0.49$ & $0.57$ \\
        \hline \hline
    \end{tabular*}
\end{table}

\begin{figure}[H]
    \centering
    \includegraphics[width=0.73\textwidth]{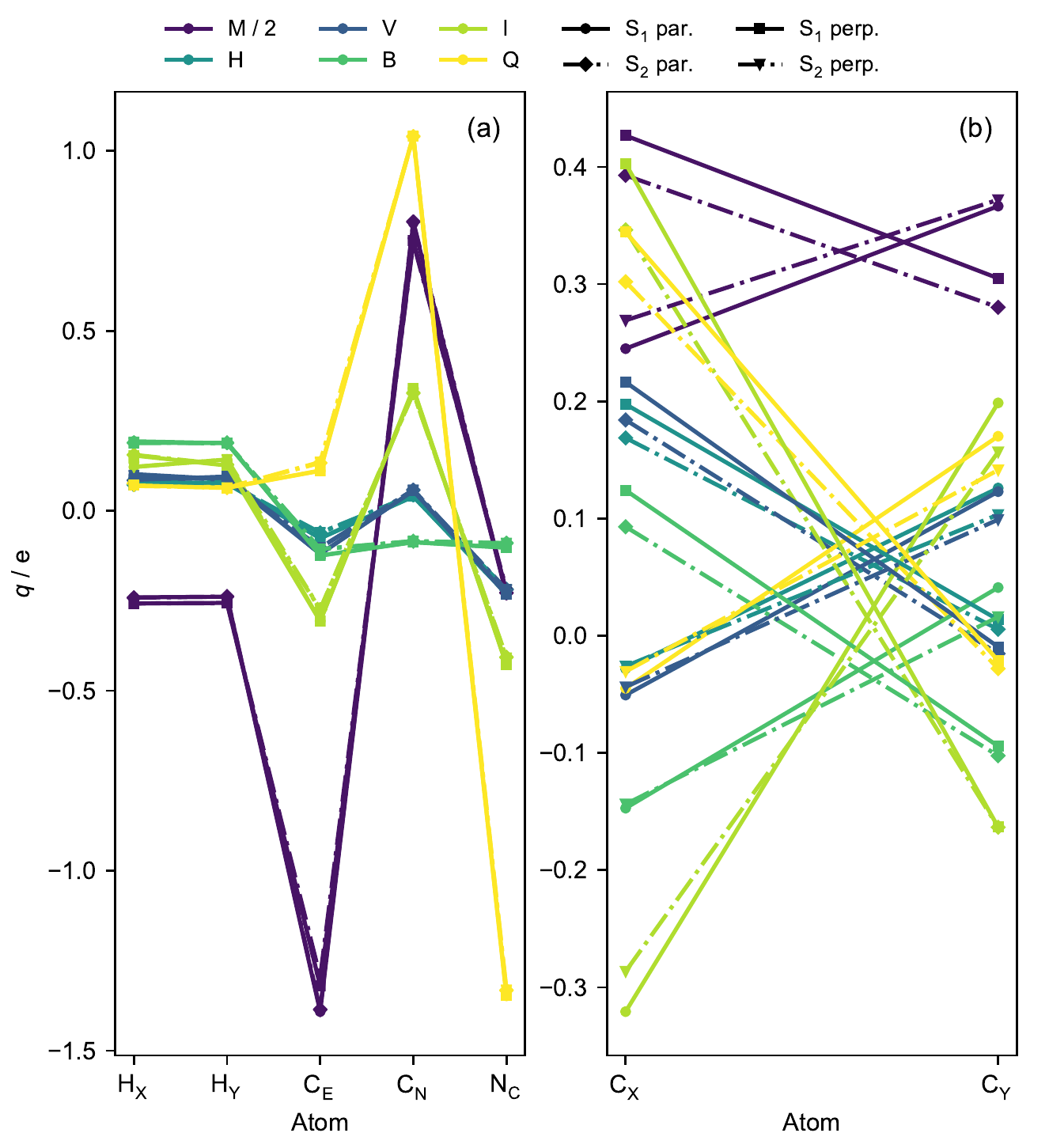}
    \caption{CASSCF charges of symmetry-equivalent (a) benzene H atoms and all TCNE atoms (b) benzene C atoms for the charge-transfer states of B-TCNE. Charge labels
    are M: Mulliken; H: Hirshfeld; V: VDD; B: Becke; I: IH; and Q: QTAIM. Mulliken charges were scaled by a factor of 2 to fit on the same scale.}
    \label{fig:btmcchg}
\end{figure}

\section{Ethylene potential energies}
\begin{figure}[H]
    \centering
    \includegraphics[width=0.5\textwidth]{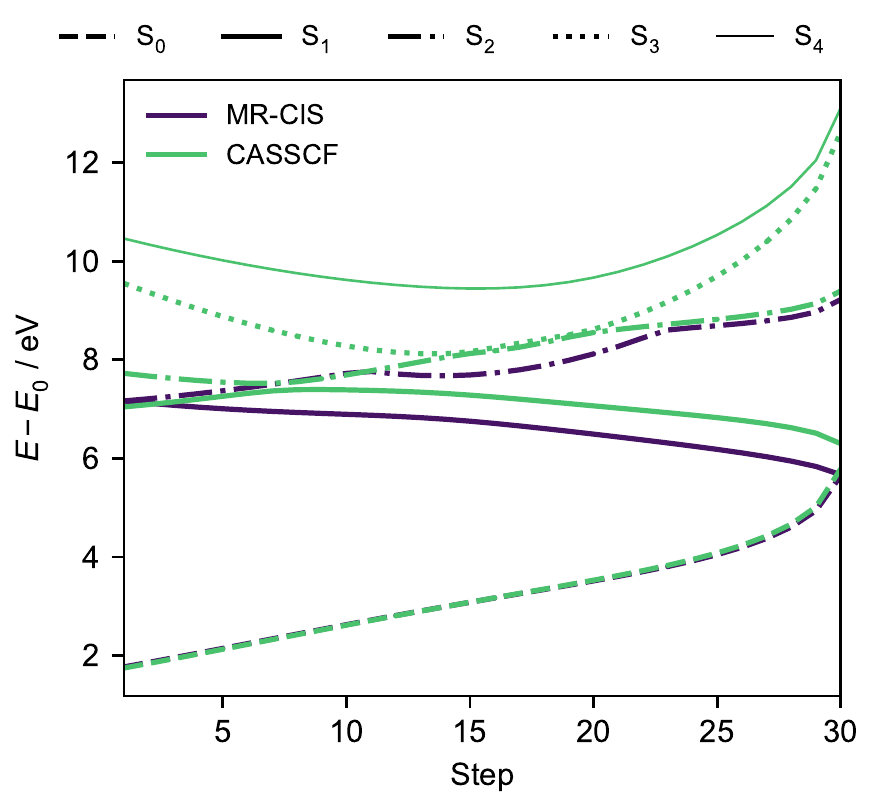}
    \caption{Potential energies along the interpolated path between S$_2$-S$_1$ and S$_1$-S$_0$ MECIs of ethylene.}
    \label{fig:etpathpes}
\end{figure}

\section{Ethylene CASSCF charges}
\begin{figure}[H]
    \centering
    \includegraphics[width=0.5\textwidth]{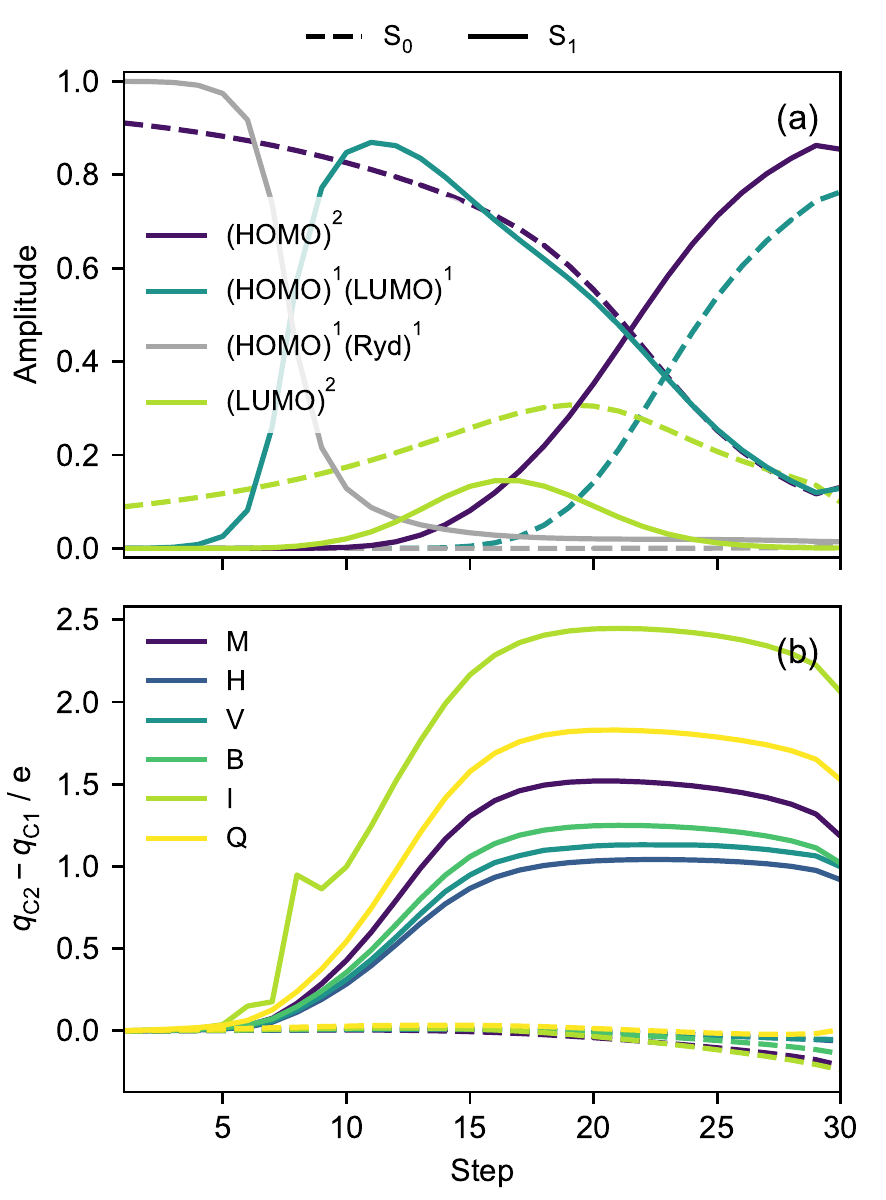}
    \caption{(a) Electronic character and (b) charge differences along interpolated internal coordinates from the S$_2$-S$_1$ twisted MECI to
    the \textbf{C1Pyr} S$_1$-S$_0$ MECI at the CASSCF level of theory. Charges in (b) are M: Mulliken; H: Hirshfeld;
    V: VDD; B: Becke; I: IH; and Q: QTAIM. Solid lines represent the ground state (S$_0$) and dashed lines represent
    the excited state (S$_1$) in (a) and (b).}
    \label{fig:etmcpath}
\end{figure}

\section{Ethylene QTAIM details}
\begin{figure}[H]
    \centering
    \includegraphics[width=\textwidth]{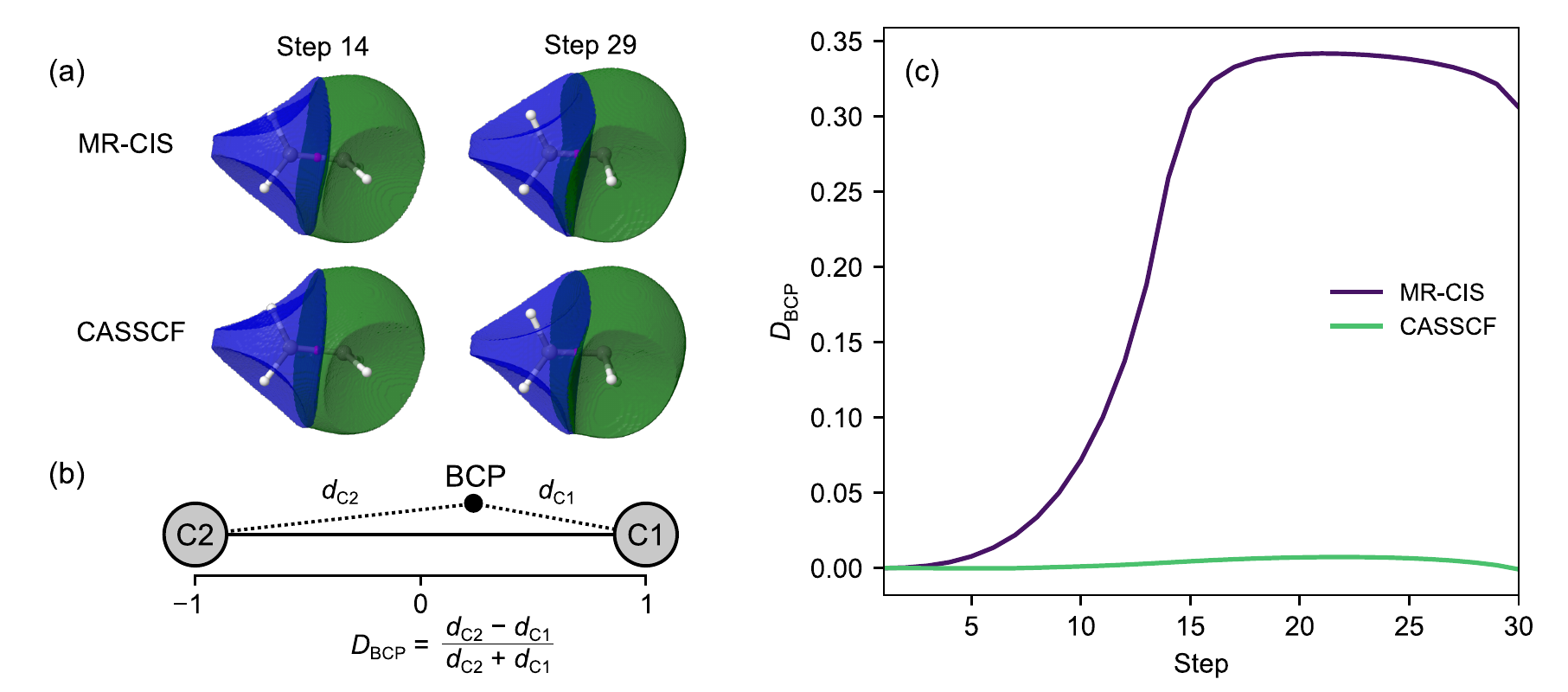}
    \caption{(a) QTAIM basins of C1 (green) and C2 (blue) at select points of the interpolated path, showing the distortion of the MR-CIS basin
    toward C1. (b) Definition of the distance metric for the bond-critical point (BCP) between C1 and C2. (c) The BCP distance metric along the
    interpolated path for MR-CIS and CASSCF. The BCP was within 3\% of the C1--C2 bond length from the C1--C2 bond for all points on the path.}
    \label{fig:etqtaim}
\end{figure}

\section{Polyene electronic structure details} \label{sec:lot}
\begin{table}[H]
    \caption{Basis set, active space and CASSCF state-averaging for MR-CIS calculations of polyenes. ET: ethylene; BD: 1,3-butadiene; HT: 1,3,5-hexatriene;
    CHD: 1,3-cyclohexadiene; Am: amino; CN: cyano; 3s: diffuse s-type Rydberg orbital; CAS: complete active space.}
    \label{tab:lot}
    \begin{tabular*}{\textwidth}{l@{\extracolsep{\fill}}cccc}
        \hline \hline
        Molecule & Basis set & CAS electrons & CAS orbitals & Number of states \\ \hline
        ET & 6-31G* + 3s(X) & 2 & 3 & 5 \\
        VAm & 6-31G* + 3s(N) & 4 & 4 & 8 \\
        VCN & 6-31G* & 6 & 6 & 5 \\ \hline
        BD & 6-31G* & 4 & 4 & 3 \\
        1-AmBD & 6-31G* & 4 & 4 & 3 \\
        2-AmBD & 6-31G* & 4 & 4 & 3 \\
        1-CNBD & 6-31G* & 6 & 6 & 5 \\
        2-CNBD & 6-31G* & 6 & 6 & 5 \\ \hline
        HT & 6-31G* & 6 & 6 & 3 \\
        1-AmHT & 6-31G* & 6 & 6 & 3 \\
        2-AmHT & 6-31G* & 6 & 6 & 3 \\
        3-AmHT & 6-31G* & 6 & 6 & 3 \\
        1-CNHT & 6-31G* & 6 & 6 & 3 \\
        2-CNHT & 6-31G* & 6 & 6 & 3 \\
        3-CNHT & 6-31G* & 6 & 6 & 3 \\
        \hline \hline
    \end{tabular*}
\end{table}

\noindent
3s(X) definition (only basis function on X for ET):
{\setstretch{1.0}
\begin{verbatim}
 S   8   1
    0.02462393         0.46870000
    0.01125334         0.71170000
    0.00585838        -0.40270000
    0.00334597         0.55780000
    0.00204842        -0.62380000
    0.00132364         0.51860000
    0.00089310        -0.27390000
    0.00062431         0.06710000
\end{verbatim}
}

\noindent
3s(N) definition (added to the 6-31G* basis functions on N for VAm):
{\setstretch{1.0}
\begin{verbatim}
 S   8   1
    0.02462393         0.42720000
    0.01125334         0.79880000
    0.00585838        -0.55920000
    0.00334597         0.83870000
    0.00204842        -0.97890000
    0.00132364         0.83340000
    0.00089310        -0.44630000
    0.00062431         0.11030000
\end{verbatim}
}

\section{Polyene MECI potential energies}
\begin{figure}[H]
    \centering
    \includegraphics[width=\textwidth]{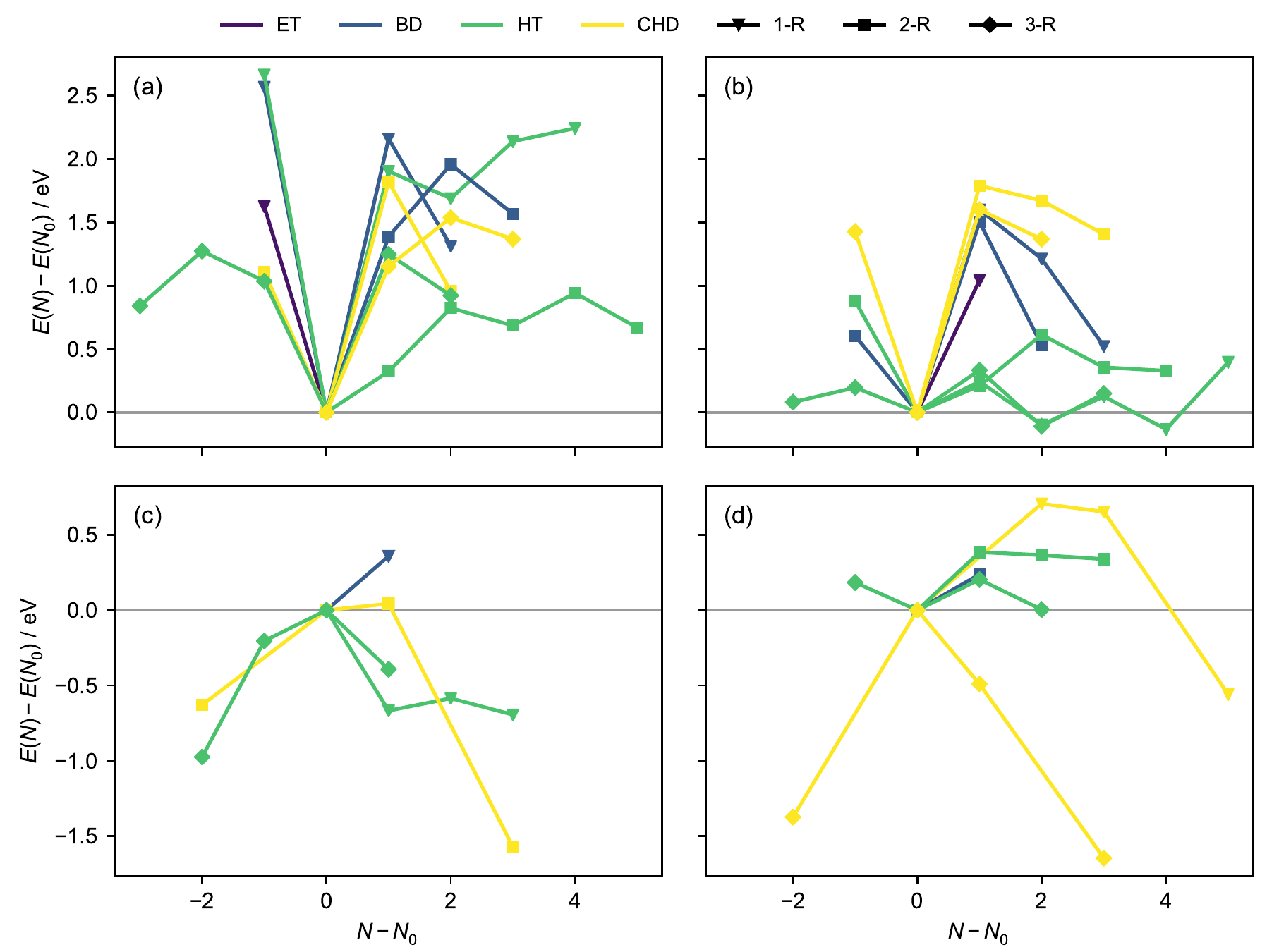}
    \caption{MR-CIS relative potential energies of (a) amino-substituted ethylenic MECIs, (b) cyano-substituted ethyleneic MECIs,
    (c) amino-substituted kinked-diene MECIs and (d) cyano-substituted kinked-diene MECIs. Energies and carbon backbone indices are given relative to
    MECI favoured to have a negative charge by the substituent, $N_0$, given by the substituted position for cyano substitution and by the carbon
    doubly-bonded to the substituted carbon for amino substitution. Only species with a valid $N_0$ are shown.}
    \label{fig:polymeci}
\end{figure}

\section{Polyene properties}
Geometries, nonadiabatic coupling vectors, gradient difference vectors, charges and energies of optimized ground-state minima and MECIs
are organized in a dataset on \\ \texttt{https://github.com/ryjmacdonell/polyene-meci-dataset}.